\documentclass[a4paper,fleqn,usenatbib,useAMS]{mnras}

\usepackage{graphicx}	
\usepackage{amsmath}	
\usepackage{amssymb}	
\usepackage{multicol}        
\usepackage{bm}		
\usepackage{pdflscape}	
\usepackage{lineno}
\usepackage{eso-pic}
\AddToShipoutPictureBG*{%
  \AtPageUpperLeft{%
    \hspace{0.75\paperwidth}%
    \raisebox{-3.5\baselineskip}{%
     \makebox[0pt][l]{\textnormal{DES 2017-0238}}
}}}%

\AddToShipoutPictureBG*{%
  \AtPageUpperLeft{%
    \hspace{0.75\paperwidth}%
    \raisebox{-4.5\baselineskip}{%
      \makebox[0pt][l]{\textnormal{Fermilab PUB-17-349-AE}}
}}}%

\usepackage[T1]{fontenc}
\usepackage{ae,aecompl}

\title[DES\,3 and DES\,J0222.7$-$5217]{Deep SOAR follow-up photometry of two Milky Way outer-halo companions discovered with Dark Energy Survey}

\author[E. Luque et al.]{E.~Luque,$^{1,2}$\thanks{\textbf{E-mail}: elmer.luque@ufrgs.br} B.~Santiago,$^{1,2}$ A.~Pieres,$^{1,2}$ J.~L.~Marshall,$^{3}$ A. B.~Pace,$^{3}$ 
\newauthor  
R.~Kron,$^{4,5}$ A.~Drlica-Wagner,$^{4}$ A.~Queiroz,$^{1,2}$ E.~Balbinot,$^{6}$ M.~dal Ponte,$^{1,2}$   
\newauthor
A.~Fausti Neto,$^{2}$ L.~N.~da Costa,$^{2,7}$ M.~A.~G.~Maia,$^{2,7}$ A.~R.~Walker,$^{8}$   
\newauthor
F.~B.~Abdalla,$^{9,10}$ S.~Allam,$^{4}$ J.~Annis,$^{4}$ K.~Bechtol,$^{11}$ A.~Benoit-L{\'e}vy,$^{9,12,13}$
\newauthor
E.~Bertin,$^{12,13}$ D.~Brooks,$^{9}$  A.~Carnero~Rosell,$^{2,7}$ M.~Carrasco~Kind,$^{14,15}$
\newauthor
J.~Carretero,$^{16}$ M.~Crocce,$^{17}$ C.~Davis,$^{18}$ P.~Doel,$^{9}$ T.~F.~Eifler,$^{19,20}$
\newauthor
B.~Flaugher,$^{4}$ J.~Garc\'ia-Bellido,$^{21}$ D.~W.~Gerdes,$^{22,23}$ D.~Gruen,$^{18,24}$
\newauthor
R.~A.~Gruendl,$^{14,15}$ G.~Gutierrez,$^{4}$ K.~Honscheid,$^{25,26}$ D.~J.~James,$^{27}$
\newauthor
K.~Kuehn,${^{28}}$ N.~Kuropatkin,$^{4}$ R.~Miquel,$^{16,29}$ R.~C.~Nichol,$^{30}$ A.~A.~Plazas,$^{20}$
\newauthor
E.~Sanchez,$^{31}$ V.~Scarpine,$^{4}$ R.~Schindler,$^{24}$ I.~Sevilla-Noarbe,$^{31}$ M.~Smith,$^{32}$
\newauthor
M.~Soares-Santos,$^{4}$ F.~Sobreira,$^{2,33}$  E.~Suchyta,$^{34}$ G.~Tarle$^{23}$ and D.~Thomas$^{30}$
\vspace*{1em}\\\noindent
Affiliations are listed at the end of the paper
}

\date{Released \today}

\pubyear{2015}

\begin{document}
\label{firstpage}
\pagerange{\pageref{firstpage}--\pageref{lastpage}}
\maketitle

\begin{abstract}
We report the discovery of a new star cluster, DES\,3, in the constellation of Indus, and deeper observations of the previously identified satellite DES\,J0222.7$-$5217 (Eridanus\,III). DES\,3 was detected as a stellar overdensity in first-year Dark Energy Survey data, and confirmed with deeper photometry from the 4.1 metre Southern Astrophysical Research (SOAR) telescope. The new system was detected with a relatively high significance and appears in the DES images as a compact concentration of faint blue point sources. We determine that DES\,3 is located at a heliocentric distance of $\sim 76\,\mathrm{kpc}$ and it is dominated by an old ($\simeq 9.8\,\mathrm{Gyr}$) and metal-poor ($\mathrm{[Fe/H]}\simeq -1.88$) population. While the age and metallicity values of DES\,3 are similar to globular clusters, its half-light radius ($r_\mathrm{h}\sim 6.5\,\mathrm{pc}$) and luminosity ($M_V \sim -1.9$) are more indicative of faint star clusters. Based on the apparent angular size, DES\,3, with a value of $r_\mathrm{h}\sim 0\farcm 3$, is among the smallest faint star clusters known to date. Furthermore, using deeper imaging of DES\,J0222.7$-$5217 taken with the SOAR telescope, we update structural parameters and perform the first isochrone modeling. Our analysis yields the first age ($\simeq 12.6\,\mathrm{Gyr}$) and metallicity ($\mathrm{[Fe/H]}\simeq -2.01$) estimates for this object. The half-light radius ($r_\mathrm{h}\sim 10.5\,\mathrm{pc}$) and luminosity ($M_V\sim -2.7$) of DES\,J0222.7$-$5217 suggest that it is likely a faint star cluster. The discovery of DES\,3 indicates that the census of stellar systems in the Milky Way is still far from complete, and demonstrates the power of modern wide-field imaging surveys to improve our knowledge of the Galaxy's satellite population.
\end{abstract}





\section{Introduction}
A fundamental prediction of the Lambda cold dark matter ($\mathrm{\Lambda CDM}$) theory of structure formation is that galactic DM haloes of the size of the Milky Way (MW) grow by the accretion of smaller sub-systems \citep[e.g.][]{White1978,Davis1985,Font2011}. The Sagittarius dwarf galaxy and the globular cluster Palomar\,5 are interesting examples of physical systems that are even now going through the process of tidal disruption before being absorbed by the MW \citep[see e.g.][]{Ibata1994,Odenkirchen2002}. Spectroscopic observations have shown that the dwarf galaxies are DM dominated systems, while there is virtually no evidence of DM halos surrounding the globular clusters \citep[e.g.][]{Willman2012,Ibata2013}.

Based on the horizontal branch (HB) morphology, metallicity, structure and kinematics, the globular clusters of the MW halo have been classified into two groups: the Young and Old Halo globular clusters \citep{Zinn1985,Zinn1993,Mackey2004,Mackey2005,Milone2014,Marino2014,Marino2015}. It is established observationally that the accretion of dwarf galaxies leads to the accretion of globular clusters, the so-called Young Halo clusters, and possibly open clusters \citep[e.g.][and references therein]{Mackey2004,Mackey2005,Carraro2009,Law2010}.  Therefore, searching for star clusters can help us understand the assembly history of our Galaxy and the associated globular cluster system.

With the advent of the Sloan Digital Sky Survey \citep[SDSS;][]{York2000}, a new class of star clusters was discovered \citep[e.g.][]{Koposov2007,Belokurov2010,Fadely2011,Munoz2012,Kim2015a}. These star clusters have very low luminosities ($-3.0\lesssim M_V\lesssim 0 $), small half-light radii ($r_\mathrm{h} < 10\,\mathrm{pc}$), and are thought to suffer mass-loss via dynamical processes such as tidal disruption or evaporation \cite[see e.g.][]{Koposov2007,Kim2015a}. The census of MW satellite galaxies has also increased considerably, from 11 classical dwarfs known in 1990, up to a total of 27 that were known by early 2015 \citep{McConnachie2012}. Over the past two years many satellite candidates have been found in the following surveys: the Dark Energy Survey (DES; \citealt{DES2005}), the Panoramic Survey Telescope and Rapid Response System 1 \citep{Laevens2014,Laevens2015a,Laevens2015b}, the Survey of the Magellanic Stellar History \citep{Martin2015a}, VST ATLAS \citep{Torrealba2016a,Torrealba2016b}, the Hyper Suprime-Cam Subaru Strategic Program \citep{Homma2016,Homma2017}, and the Magellanic SatelLites Survey \citep{Drlica2016}. In particular, 21 stellar system candidates with $M_V \gtrsim -8$ have been found in DES  \citep{Bechtol2015,Drlica2015,Koposov2015,Kim2015b,Luque2016,Luque2017}. Thus far, spectroscopic measurements of radial velocity and metallicity have confirmed that Reticulum II \citep{Koposov2015b,Simon2015,Walker2015}, Horologium I \citep{Koposov2015b}, Tucana II \citep{Walker2016}, Grus I \citep{Walker2016}, Tucana III \citep{Simon2017}, and Eridanus\,II \citep{Li2017} are indeed dwarf galaxies. The possible association of the recently discovered DES dwarf galaxy candidates with Large and Small Magellanic Clouds (LMC and SMC) has been discussed by several authors \citep[e.g.][]{Bechtol2015,Drlica2015,Drlica2016,Koposov2015,Jethwa2016,Dooley2017,Sales2017}. Several of the candidates may be associated with the Sagittarius stream \citep{Luque2017}.

Here, we announce the discovery of a new MW star cluster, which we call DES\,3, in the constellation of Indus. The object was detected as a stellar overdensity in the first internal release of the DES co-add data (Y1A1), which covers a solid angle of $\sim 1800 \deg^2$ in the southern equatorial hemisphere, and later confirmed with deep SOAR imaging. We additionally present deeper imaging of DES\,J0222.7$-$5217 (Eridanus\,III) taken with the SOAR telescope in order to determine its properties, several of which have not been reported in the literature \citep{Bechtol2015,Koposov2015}. We use these deeper data to better investigate the nature of this object, whether it be a star cluster or a very small faint dwarf. If DES\,J0222.7$-$5217 is confirmed to be a star cluster, it will be named DES\,4. This paper is organized as follows. Section \ref{sec:method} describes the method used to search for star clusters and other stellar systems. In Section \ref{sec:data}, we describe the first-year DES data and discovery of DES\,3. The photometric follow-up observations and data reduction are presented in Section \ref{sec:follow-up}. In Section \ref{sec:param}, we quantify the physical properties of the new stellar object. In Section \ref{sec:DES4}, we present the updated properties of DES\,J0222.7$-$5217 with deeper imaging. Our final remarks are given in Section \ref{sec:conclusions}.

\section{substructure search method}
\label{sec:method}
Here we briefly review our overdensity search technique \citep[\textsc{sparsex};][]{Luque2016}. The {\sc sparsex} code is based on the matched-filter (MF) method \citep{Rockosi2002,Szabo2011}. It minimizes the variance between star counts from a data catalogue and a model containing a simple stellar population (SSP) and a MW field star contamination. This minimization is carried out over colour and magnitude bins for each individual and small spatial cell. A grid of SSPs covering a wide range of ages, metallicities, and distances, is created using the code {\tt\sc gencmd}\footnote{https://github.com/balbinot/gencmd}. The model of field stars is created from the catalogue data over $10\degr\times 10\degr$ regions, to account for possible variations in the field population mix. The variance minimization yields the SSP model normalization as a function of position, i.e., a map with the number density of stars consistent with that particular SSP. We then convolve each SSP map with different spatial kernels to highlight substructures of different sizes. For each of these maps, we  apply \textsc{SExtractor} \citep{Bertin1996} to identify the most significant and/or frequently detected overdensities. Then we perform visual inspection on the images, colour-magnitude diagram (CMD) and significance profile of each candidate. The significance profile is defined as the ratio of the number of stars $N_\mathrm{obj}$ inside a given radius in excess of the number $N_\mathrm{bgd}$, relative to the expected fluctuation in the same field background, i.e, $N_\mathrm{obj}/\sqrt{N_\mathrm{bgd}}$. Defining $N_\mathrm{obs}$ to be the total number of observed stars, then $N_\mathrm{obj}=(N_\mathrm{obs}- N_\mathrm{bgd})$. To avoid a low stellar statistic, we built the significance profile using a cumulative radius of $1\arcmin$ centred on the candidate. $N_\mathrm{bgd}$ is computed within a circular annulus at $30\arcmin < r < 36\arcmin$ from each candidate. We refer to \cite{Luque2016, Luque2017} and to Sections \ref{sec:data} and \ref{sec:param} below for more details.

\section{DES Data and discovery}
\label{sec:data}
DES is a wide-field optical survey that uses the Dark Energy Camera (DECam; \citealt{Flaugher2015}) to image $5000\,\deg^2$ in the southern equatorial hemisphere. DECam is an array of $62$ $\mathrm{2k\times 4k}$ CCDs, with pixel scale of $0\farcs 263$, that images a $2\fdg 2$ diameter field of view. It is installed at the prime focus of the $4$-metre Blanco telescope at Cerro Tololo Inter-American Observatory.  DECam images are reduced by the DES Data Management (DESDM) system. The pipeline consists of image detrending, astrometric calibration, nightly photometric calibration, global calibration, image coaddition, and object catalogue creation \citep[see][for a more detailed description]{Sevilla2011,Desai2012,Mohr2012,Balbinot2015,Drlica-Wagner2017}. The \textsc{SExtractor} toolkit is used to create catalogues from the processed and co-added images \citep{Bertin1996,Bertin2011}. 

\begin{figure}\centering
\includegraphics[width=.45\textwidth]{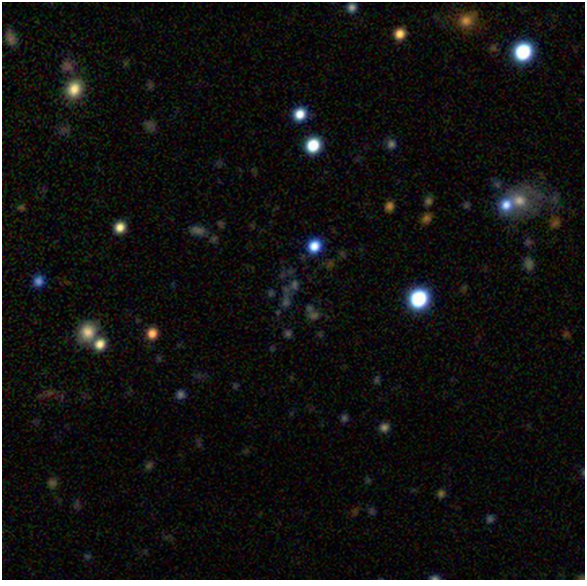}
\caption{DES co-add image cutout of DES\,3 taken from the DES Science portal. The $1.58\, \mathrm{arcmin}\times 1.58\, \mathrm{arcmin}$ image is centred on DES\,3. The R,G,B channels correspond to the $i$, $r$, $g$ bands.}\label{fig:DES3}
\end{figure}

To search for stellar substructures in the DES Y1A1 catalogue, we applied cuts based on the  \textsc{SExtractor} parameters $\mathtt{SPREAD\_MODEL}$, $\mathtt{FLAGS}$ and point spread function ($\mathtt{PSF}$) magnitudes. The \texttt{FLAGS} parameter denotes if an object is saturated or has been truncated at the edge of the image and a cut of $\mathtt{FLAGS}<4$ is sufficient for our purposes. We apply this restriction in all our subsequent analyses. The $\mathtt{SPREAD\_MODEL}$ parameter is the main star/galaxy separator. To avoid issues arising from fitting the \texttt{PSF} across variable-depth co-added images, we utilized the weighted-average ($\mathtt{WAVG}$) of the $\mathtt{SPREAD\_MODEL}$ measurements from the single-epoch exposures \citep[see][]{Bechtol2015}. Therefore, our stellar sample consists of sources in the $i$ band with $|\mathtt{WAVG\_SPREAD\_MODEL}|<0.003+\mathtt{SPREADERR\_MODEL}$ as described in \citet{Drlica2015} and \citet{Luque2016}. In addition, magnitude\footnote{We refer the $\mathtt{WAVG\_MAG\_PSF}$ measurements in the DES $gri$ filters as $g_\mathrm{DES}$, $r_\mathrm{DES}$ and $i_\mathrm{DES}$, respectively.} ($17<g_\mathrm{DES}<24$) and colour ($-0.5<g_\mathrm{DES}-r_\mathrm{DES}<1.2$) cuts were also applied. The colour cut was performed to exclude stars from the Galactic disc and possibly spurious objects that can contaminate our sample. Each star was extinction corrected from the reddening map of \cite{Schlegel1998}.

Applying the method described in Section \ref{sec:method} on DES Y1A1 data, we have successfully recovered all ten stellar objects that have been reported in first-year DES data\footnote{The eleventh object, Grus\,1, reported by \citet{Koposov2015} is in a region of Y1 data that is not included in the Y1A1 coadd due to limited coverage in some of the DES filters.} \citep{Bechtol2015, Koposov2015,Kim2015b,Luque2016}. These detections include the faint satellite DES\,J0222.7$-$5217, which we detect with significance of $16.1\sigma$. \citet{Bechtol2015} reported DES\,J0222.7$-$5217 with a heliocentric distance of $95\,\mathrm{kpc}$, an absolute magnitude of $-2.4\pm 0.6$, and a half-light radius of $11^{+8}_{-5}\,\mathrm{pc}$. In contrast, \citet{Koposov2015} found DES\,J0222.7$-$5217 to be slightly less distant, $87\,\mathrm{kpc}$, with an absolute magnitude of $-2.0\pm 0.3$, and a half-light radius of $14.0^{+12.5}_{-2.6}\,\mathrm{pc}$. Deeper imaging is necessary to conclusively  determine the characteristics of this faint satellite.

\begin{figure*}\centering
\includegraphics[width=1.0\textwidth]{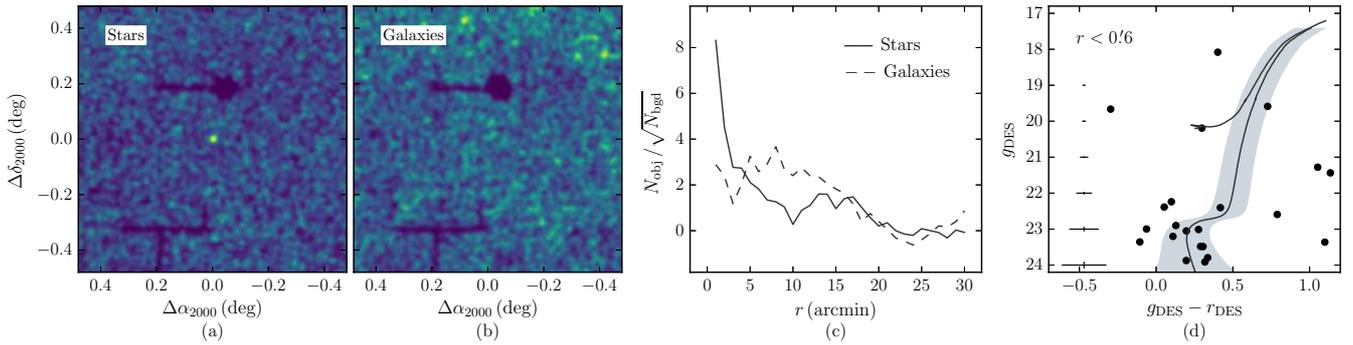}
\caption{Detection of DES\,3 from the DES data. Panel (a): stellar density map around DES\,3. Panel (b): similar to previous panel, but now for galaxies. Panel (c): significance profile as a function of radius $r$ from the centre of DES\,3. The solid line corresponds to stars, while the dashed line corresponds to galaxies. Panel (d): CMD of stars within a circle with radius $r=0\farcm 6$ from the centre of DES\,3. A \textsc{parsec} (solid line) isochrone model with age $9.8\,\mathrm{Gyr}$ and  $\mathrm{[Fe/H]}=-1.88$ is overplotted at a distance of $76.2\,\mathrm{kpc}$ (see Section \ref{sec:param} for details of the best-fitting isochrone). The isochrone filter (gray shaded area) based on photometric uncertainties contains the most likely members. The mean photometric errors in both colour and magnitude are shown in the extreme left of this panel.}\label{fig:mapden}
\end{figure*}

In addition, to the previously detected satellites, we detected one new star cluster, DES\,3, with a statistical significance of $8.3\sigma$. DES\,3 is readily visible as a cluster of faint blue stars in the DES co-add images\footnote{The images were taken from the DES Science portal. The latter is a web-based system being developed by DES-Brazil and Laborat\'orio Interinstitucional de e-Astronomia (LIneA, http://www.linea.gov.br) for the DES collaboration.} (see Fig. \ref{fig:DES3}). The fact that they are blue sources indicates that this overdensity is not a galaxy cluster. The overdensity is also clear from the DES stellar map shown in panel (a) of Fig. \ref{fig:mapden}. For comparison, in panel (b), we show the density map of sources classified as galaxies. Note that faint galaxies do not contribute to the observed overdensity of DES\,3. In panel (c) of Fig. \ref{fig:mapden}, the circular significance profile is shown.  Note that the significance profile shows the higher peak at $r=1\arcmin$ from the centre of DES\,3, where $N_\mathrm{obj}=22$ and $N_\mathrm{bgd}=7$ stars. To take into account the missing coverage\footnote{We masked the regions on the sky where there is absence of sources.} observed in panels (a) and (b) of Fig. \ref{fig:mapden}, in this analysis, we estimated the effective area ($A_\mathrm{eff}$) for each circular region as follows. After inserting a certain number of uniformly distributed random points ($N$) inside a circle with radius $r$ centred on DES\,3, we computed the ratio between the number of points that fall inside the region of the sky that is covered by the DES data ($N_\mathrm{in}$) and $N$. Therefore, for each region $A_\mathrm{eff}=A\frac{N_\mathrm{in}}{N}$, where $A$ is the area of the circle.

In panel (d) of Fig. \ref{fig:mapden}, we show the CMD of DES\,3 constructed with stars within a circle of radius $r=0\farcm 6$. We also have plotted a \textsc{parsec} \citep[CMD v3.0;\footnote{We are using \textsc{parsec} isochrones revised to include instrumental and atmosphere response.}][]{Bressan2012} isochrone model (solid line) corresponding to the best-fitting parameters (this will be discussed fully in Section \ref{sec:param}). There are stars with $g_\mathrm{DES} \gtrsim 22.5\,\mathrm{mag}$ scattered in the CMD, some of which fall inside the isochrone filter\footnote{The isochrone filter is built by using the photometric uncertainties in both colour and magnitude. We added a value of $0.1\,\mathrm{mag}$ in the colour-magnitude space to avoid too narrow isochrone filters at the bright magnitudes, where the uncertainties are small \citep[for details, see ][]{Luque2016}.} (gray shaded area). We are tempted to say that these stars belong to the main sequence turn-off (MSTO) and sub-giant branch (SGB) of DES\,3. Note that the CMD also shows one star that may belong to the HB. However, based on the limited information given by this CMD, we are not able to confirm the nature and infer reliable parameters for DES\,3. In order to reach the main sequence (MS; $g_\mathrm{DES}\sim 25.5\,\mathrm{mag}$) in the CMD, deeper follow-up observations are required.

\section{SOAR follow-up data}
\label{sec:follow-up}
Follow-up imaging of DES\,3 and DES\,J0222.7$-$5217 was carried out on 2016 July 29 and October 20 respectively, using the SOAR Optical Imager (SOI) on the $4.1$-metre Southern Astrophysical Research (SOAR) telescope. SOI consists of two $2048\times 4096$ CCDs and covers a $5\farcm 2\times 5\farcm 2$ field. The SOI CCDs have a scale of $0\farcs 077/\mathrm{pixel}$. As the images were binned $2\times 2$, the final image scale is $0\farcs 154/\mathrm{pixel}$. We observed each object for a total of 45\,min in each SDSS filter ($g$ and $r$; hereafter $g^\prime$ and $r^\prime$). The integrations were split into nine exposures of $300\,s$ to avoid overexposing. The observations were carried out with an airmass below $1.20$.   

\begin{figure}\centering
\includegraphics[width=.45\textwidth]{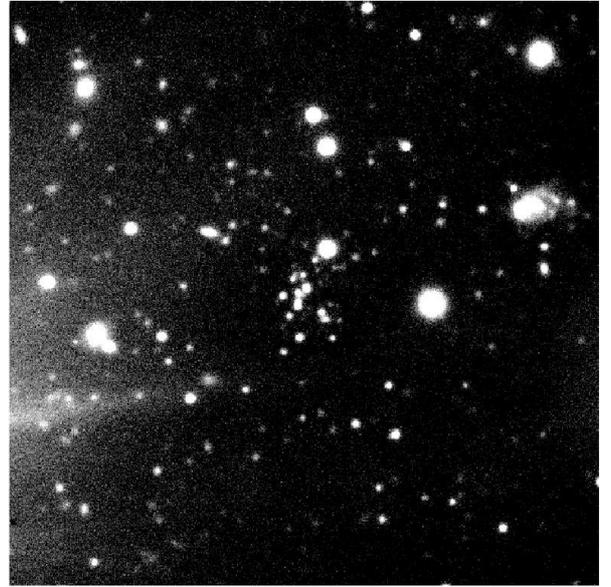}
\caption{SOAR $g$ band co-add image cutout of DES\,3. The $1.58\, \mathrm{arcmin}\times 1.58\, \mathrm{arcmin}$ image is centred on DES\,3.}\label{fig:DES3_SOAR}
\end{figure}

Raw exposures were trimmed, corrected for bias, and flat fielded by the SOAR Brazilian Resident Astronomers, David Sanmartin and Luciano Fraga, using \textsc{soar/iraf} packages. Individual exposures for each filter, $g^\prime$ and $r^\prime$, were co-added. In particular, Fig. \ref{fig:DES3_SOAR} shows the SOAR/SOI ($g^\prime$ band only) coadded images of DES\,3. The figure attests to the increase in spatial resolution\footnote{By using several bright and isolated stars in the direction of DES\,3, we determined that the DES co-added images have an average $\mathtt{PSF}$ FWHM of $\simeq 1\farcs 19$ and $0\farcs 94$ in the $g$ and $r$ bands, respectively, while the SOAR co-added images have a value of $\simeq 0\farcs 8$ in both $g^{\prime}$ and $r^{\prime}$ bands.} and in photometric depth when the latter is compared with the DES images (see Fig. \ref{fig:DES3}). The scattered light observed in Fig. \ref{fig:DES3_SOAR} is discussed in the text below. To create a source catalogue in the direction of DES\,3 and DES\,J0222.7$-$5217, we use a combination of the \textsc{SExtractor/PSFex} routines \citep{Bertin1996,Bertin2011}. A first pass of \textsc{SExtractor} is run to create an input catalogue for \textsc{PSFex}. The \textsc{PSFex} routine creates a $\mathtt{PSF}$ image for a second pass of \textsc{SExtractor} that determines the $\mathtt{PSF}$ magnitude and the $\mathtt{SPREAD\_MODEL}$ parameter.

We transformed the instrumental magnitudes, $g_\mathrm{inst}$ and $r_\mathrm{inst}$, to apparent magnitudes\footnote{The $g_\mathrm{inst}$ and $r_\mathrm{inst}$ magnitudes are given by $-2.5\times\log(\mathrm{counts})$, while the $g$ and $r$ magnitudes correspond to $g_\mathrm{DES}$ and $r_\mathrm{DES}$, respectively.}, $g$ and $r$, by using a set of DES stars in the direction of each object. This process was performed as follows: first we built a SOAR catalogue by merging the $g^\prime$ and $r^\prime$ full photometric lists, using a matching radius of $0\farcs 5$.  We also selected DES stellar sources with ${\tt |WAVG\_SPREAD\_MODEL|} < 0.003$ and ${\tt WAVG\_MAGERR\_PSF} < 0.03$ to obtain a sufficiently pure stellar sample. Next, we matched the SOAR sources with DES brightest (from $17$ to $22\,\mathrm{mag}$) stars, on $g^\prime$ and $r^\prime$ filters, using a radial tolerance of $0\farcs 5$. Then, we use these brightest stars to fit the following calibration curves:
\begin{align}
\begin{aligned}
g &= g_\mathrm{inst}+\beta+\gamma(g_\mathrm{DES}-r_\mathrm{DES}),\\
r &= r_\mathrm{inst}+\zeta+\eta(g_\mathrm{DES}-r_\mathrm{DES}),
\end{aligned}\label{eq:1}
\end{align}
where the $\beta$ and $\zeta$ coefficients represent the zero points of the two bands, while $\gamma$ and $\eta$ are the colour term coefficients. Airmass correction, in both filters, was assumed to be constant and, therefore, it is absorbed by the zero point coefficient. The results of our fits are presented in Table \ref{tab:calib}. Estimates for the two objects are identical within uncertainties. The calibration was then applied to all instrumental magnitudes in the photometric list. Finally, all magnitudes of the calibrated sources were corrected for Galactic reddening using the \citet{Schlegel1998} dust maps.

For our further analysis, we modify the cuts to reach faint magnitudes. Our stellar sample consists of sources with $|\mathtt{SPREAD\_MODEL}| < 0.003 + \mathtt{SPREADERR\_MODEL}$.  Moreover, we applied  colour ($-0.5<g_\mathrm{DES}-r_\mathrm{DES} < 1.2$) and magnitude ($g_\mathrm{DES}>17$) cuts. The faint $g_\mathrm{DES}$ magnitude, for each catalogue, is established based on the magnitude error ($\sigma_g$), where $\sigma_g$ has a value of $\approx 0.07\,\mathrm{mag}$. Therefore, this limiting magnitude corresponds to $g_\mathrm{DES} < 25$ ($g_\mathrm{DES}<24.5$) for the star catalogue in the direction of DES\,3 (DES\,J0222.7$-$5217).

We have verified that the scattered light observed only in the DES\,3 images (see Fig. \ref{fig:DES3_SOAR}) does not affect the completeness of sources (stars and galaxies) down to a magnitude depth $g_\mathrm{DES}\sim 25\,\mathrm{mag}$. To check we count sources as a function of magnitude both inside and outside scattered light affected regions. These regions have equal areas, which contain $159$ and $154$ sources, respectively. Thus, we find that both counts of sources in function of the magnitude are similar.
\begin{table}\centering
  \caption{Calibration coefficients obtained from the fit of the set of equations presented in equation \ref{eq:1}.}
  \begin{tabular}{lcc}\hline
    Coefficient & DES\,3 & DES\,J0222.7$-$5217\\\hline
    $\beta\ (\mathrm{mag})$ & $31.71\pm 0.02$ & $31.43\pm 0.08$\\
    $\gamma$& $-0.10\pm 0.02$ & $-0.16\pm 0.09$\\
    $\zeta\ (\mathrm{mag})$& $31.70\pm 0.02$ & $31.51\pm 0.03$\\
    $\eta$& $-0.15\pm 0.03$ & $-0.12\pm 0.02$\\\hline
  \end{tabular}\label{tab:calib}
\end{table}
\begin{figure}\centering
\includegraphics[width=.48\textwidth]{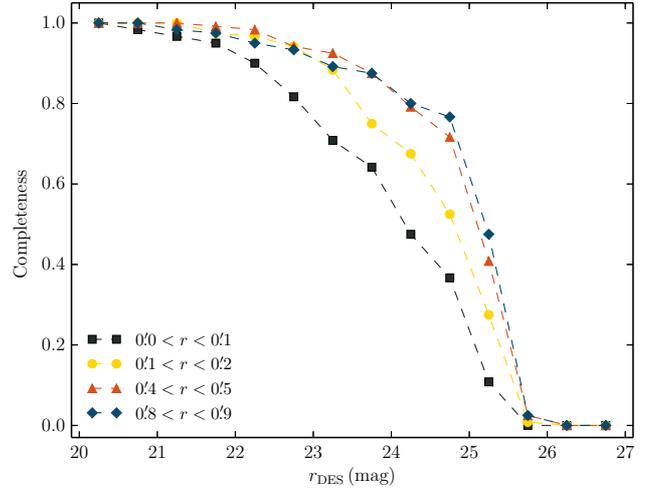}
\caption{Completeness curves as function of magnitude and radius from the SOAR data. The solid squares represent the completeness at a radius of $0\farcm 1$ from the centre of DES\,3, the solid dots represent the completeness within an annulus with radius $0\farcm 1 < r < 0\farcm 2$,  the solid triangles represent the completeness within an annulus with radius $0\farcm 4 < r < 0\farcm 5$, and the solid diamonds represent the completeness within an annulus with radius $0\farcm 8 < r < 0\farcm 9$. }\label{fig:comp}
\end{figure}

The completeness of our photometry has been evaluated performing artificial star tests on the SOAR images. The completeness curves are constructed with reference to the centre of each object up to an external radius $r=1\arcmin$, taking regular intervals of $\Delta r= 0\farcm 1$. Artificial stars from 20th to 27th magnitudes in steps of $0.5\,\mathrm{mag}$ were added to the same image with the \textsc{iraf/addstar} routine using a \texttt{PSF} model derived from several bright and isolated stars. In order to avoid the crowding in the images, only 15 per cent of the original number of sources (stars and galaxies) detected in each region were added per run\footnote{After multiple runs, we obtained a total of 120 artificial stars in each bin of magnitude.}. After inserting the artificial stars, these artificial images were reduced with the same \textsc{SExtractor/PSFex} routines as the SOAR images. \textsc{SExtractor/PSFex} assigns each detected object a $\mathtt{SPREAD\_MODEL}$ value. We consider point source candidates those sources with $|\mathtt{SPREAD\_MODEL}| < 0.003 + \mathtt{SPREADERR\_MODEL}$ (see text above for details). At last, the artificial stars are considered to be recovered if the input and output positions are closer than $0\farcs 5$, and magnitude differences are less than $0.5\,\mathrm{mag}$. The completeness curves as a function of magnitude and distance from the centre of DES\,3 are shown in Fig. \ref{fig:comp}.

\section{Properties of DES\,3}
\label{sec:param}
In order to better constrain the properties and the nature of DES\,3, we use SOAR data, which is  $\sim 1\,\mathrm{mag}$ deeper than the DES data in the region of this stellar object (see Fig. \ref{fig:DES3_SOAR}).

We use the maximum likelihood method to determine the structural and CMD parameters for DES\,3. We used the \textsc{emcee}\footnote{http://dan.iel.fm/emcee/current/} Python package \citep{Foreman2013}, which implements an affine invariant Markov Chain Monte Carlo (MCMC) ensemble sampler, to sample the $\ln$--likelihood function over the parameter space. We have assumed flat priors for all parameters. We take the median of each marginalized posterior distribution function (PDF) to be the best-fitting solution, with uncertainties given by the 16th and 84th percentiles, equivalent to $\pm 1\sigma$ assuming the PDFs are normal distributions.

\begin{figure}\centering
\includegraphics[width=.49\textwidth]{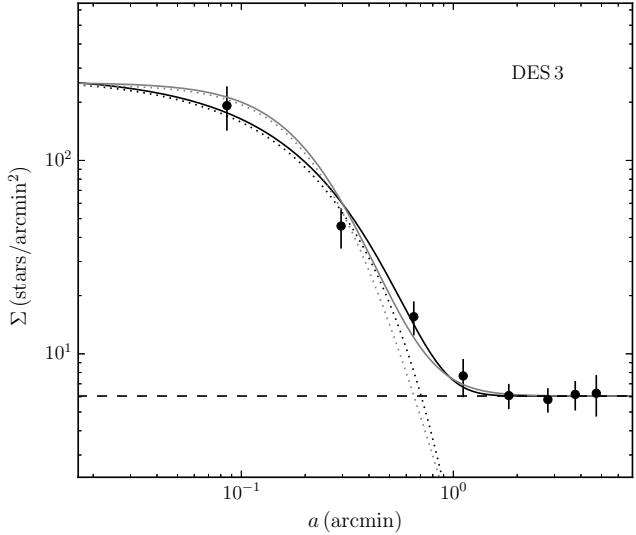}
\caption{Filled points show a binned version of the density profile of DES\,3, constructed in elliptical annuli using the derived structural parameters from the best-fitting exponential profile (see Table \ref{fitpars}). The error bars are $1\sigma$ Poisson uncertainties. The gray (black) dotted line represent the best-fitting of Plummer (exponential) profile. The horizontal dashed line shows the field background level. The gray (black) solid line is the combination of the background level with the Plummer (exponential) profile.}
\label{fig:profile}
\end{figure}

To estimate the structural parameters, we follow a convention similar to that of \citet{Martin2008}. We adopt two different density profile models: exponential and Plummer \citep{Plummer1911}. Due to the small field covered by the SOAR/SOI images, we parameterize both models with just five free parameters. For the exponential profile, the free parameters are: central coordinates $\alpha_0$ and $\delta_0$, position angle $\theta$, ellipticity $\epsilon$ and exponential scale radius $r_\mathrm{e}$. For the Plummer profile, the parameters are: $\alpha_0$, $\delta_0$,  $\theta$, $\epsilon$ and Plummer scale radius $r_\mathrm{p}$. The exponential scale radius is related to the half-light radius by the relation $r_\mathrm{h} = 1.68r_\mathrm{e}$, whereas the Plummer scale radius, $r_\mathrm{p}$, is equivalent to $r_\mathrm{h}$. The background density, $\Sigma_\mathrm{bgd}$, is determined by using a region outside $r > 1 \farcm 5$ around DES\,3, which results in $6.1\,\mathrm{\frac{stars}{arcmin^2}}$, and it is kept constant in the fits.  

\begin{table}
\caption{Properties of DES\,3.}\label{tab:DES3}
\hspace*{-.6cm}
\scalebox{.92}{
\begin{tabular}{lccc}\hline
Parameters & Exponential profile & Plummer profile & Unit\\\hline
$\alpha_0\,(J2000)$& $21\; 40\; 13.27^{+0.09}_{-0.09}$ & $21\; 40\; 13.20^{+0.11}_{-0.11}$ &$\mathrm{h\; m\; s}$\\
$\delta_0\,(J2000)$& $-52\; 32\; 31.20^{+1.50}_{-1.50}$ & $-52\; 32\; 30.48^{+1.62}_{-1.68}$ &$\mathrm{\degr\; \arcmin\; \arcsec}$\\
$l$&$343.83$&$343.83$&$\mathrm{deg}$\\
$b$&$-46.51$&$-46.51$&$\mathrm{deg}$\\
$\mathrm{D}_{\sun}$& $76.2^{+2.8}_{-4.9}$ & $76.2^{+3.2}_{-5.3}$ &$\mathrm{kpc}$\\
$r_\mathrm{h}$&$0.31^{+0.04}_{-0.03}$& $0.28^{+0.04}_{-0.03}$ &  $\mathrm{arcmin}$ \\
$r_\mathrm{h}$& $6.87^{+0.92}_{-0.80}$\textsuperscript{$a$}& $6.21^{+0.92}_{-0.79}$\textsuperscript{$a$} &$\mathrm{pc}$\\
$\theta$& $-47.0^{+31.1}_{-32.8}$ & $-34.9^{+28.8}_{-25.7}$ & $\mathrm{deg}$\\
$\epsilon$& $0.13^{+0.12}_{-0.09}$ & $0.17^{+0.13}_{-0.11}$&\\
$\Sigma_{\mathrm{bgd}}$& $6.1 \pm 0.5$ & $6.1 \pm 0.5$ & $\mathrm{\frac{stars}{arcmin^2}}$\\
$M_V$ & $-1.8^{+0.4}_{-0.3}$ & $-2.0^{+0.4}_{-0.3}$ & $\mathrm{mag}$ \\
$\mathrm{[Fe/H]}$ & $-1.88^{+0.17}_{-0.13}$ &$-1.88^{+0.22}_{-0.15}$& $\mathrm{dex}$ \\
$\mathrm{Age}$& $9.8^{+1.4}_{-1.1}$ &$9.8^{+1.4}_{-1.1}$ & $\mathrm{Gyr}$\\
$(m-M)_0$ & $19.41^{+0.08}_{-0.14}$ & $19.41^{+0.09}_{-0.15}$& $\mathrm{mag}$\\\hline
\end{tabular}}\label{fitpars}\\
\leftline{\scalebox{.91}{\textit{Note.} \textsuperscript{$a$}\footnotesize{Adopting a distance of $76.2\,\mathrm{kpc}$.}}}
\end{table}

Fig. \ref{fig:profile} shows the binned elliptical density profile of DES\,3. To determine the effective area of each elliptical annulus, correcting for the gap and borders of the field covered by the SOI CCDs, we follow the same technique used in Section \ref{sec:data}. The best-fitting exponential and Plummer models are also overplotted in this figure. As it can be seen, both the exponential and Plummer profiles adequately describe the observed data. From both models, we find that DES\,3 is only slightly elongated ($\epsilon \sim 0.15$) and compact, with a half-light radius of $r_\mathrm{h}\sim 0\farcm 3$. Its ellipticity is very similar to Kim\,2 ($\epsilon\simeq 0.12$; \citealt{Kim22015}) and Kim\,3 ($\epsilon \simeq 0.17$; \citealt{Kim32016}). Only Koposov\,1 and Koposov\,2, with $r_\mathrm{h}\sim 0\farcm 21$ and $r_\mathrm{h}\sim 0\farcm 26$, respectively \citep{Koposov2007}, have slightly smaller apparent angular sizes than DES\,3. The set of structural parameters for DES\,3 is presented in Table \ref{tab:DES3}.

\begin{figure*}\centering
\includegraphics[width=1\textwidth]{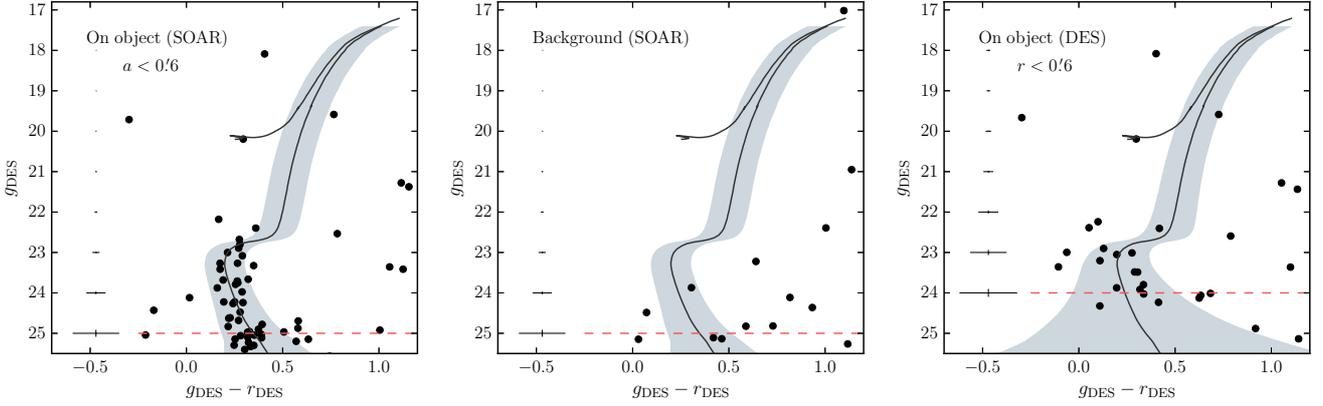}
\caption{Left panel: CMD of DES\,3 from the SOAR data. Only stars inside an ellipse with semi-major axis $a\sim 2r_\mathrm{h}$ from the centre of DES\,3 are shown. In this and the other two panels, the best-fitting \textsc{parsec} \citep{Bressan2012} isochrone  derived from the SOAR data is shown. The isochrone filter (gray shaded area) based on photometric uncertainties contains the most likely members. Middle panel: CMD of field stars in an elliptical annulus of equal area on the sky as the previous panel. Right panel: CMD of DES stars within a radius $r=0\farcm 6$ centred on DES\,3. The horizontal dashed line in each panel indicates the faint magnitude limit used. The mean photometric errors in both colour and magnitude are shown in the extreme left of each panel.}\label{fig:cmds}
\end{figure*}

For CMD fits, we first weight each star by the membership probability $p$ taken from the best profile fits. We then selected all the stars with a threshold  of $p\geqslant 0.01$ to fit an isochrone model. The free parameters age, $(m-M)_0$ and metallicity\footnote{We adopted $\mathrm{Z_\odot}=0.0152$ \citep{Bressan2012} in order to convert from  $Z$  to $\mathrm{[Fe/H]}$, assuming $\mathrm{[Fe/H]}=\log(Z/\mathrm{Z_\odot})$.}, $Z$, are simultaneously determined by this fitting method \citep{Luque2016,Luque2017,Pieres2016}. To investigate a possible range in the CMD parameters, in this analysis we use the selected stars from both exponential and Plummer models. 

\begin{figure}\centering
\includegraphics[width=.475\textwidth]{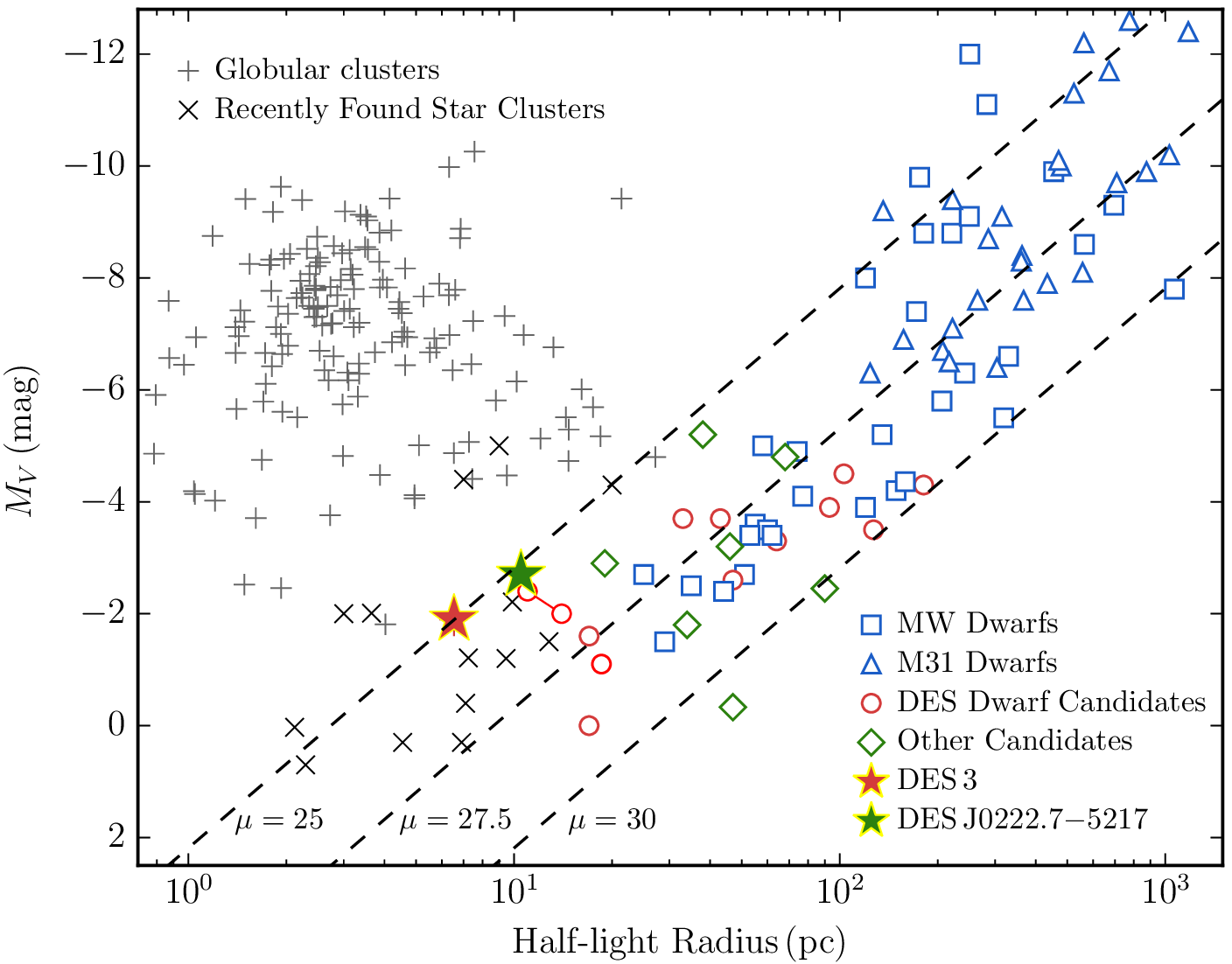}
\caption{Absolute magnitude as a function of half-light radius. MW globular clusters (`$+$' symbols; \citealt{Harris2010}), recently found MW star clusters (`$\times$' symbols; \citealt{Koposov2007}; \citealt{Belokurov2010}; \citealt{Munoz2012}; \citealt{Balbinot2013};  \citealt{Laevens2014,Laevens2015b}; \citealt{Kim2015a,Kim22015,Kim32016}; \citealt{Luque2016,Luque2017}; \citealt{Koposov2017}), MW dwarf galaxies (blue squares; \citealt{McConnachie2012}; \citealt{Bechtol2015}, \citealt{Drlica2015}; \citealt{Koposov2015}; \citealt{Kim2016b}; \citealt{Torrealba2016a,Torrealba2016b}), M\,31 dwarf galaxies (blue triangles; \citealt{McConnachie2012}), previously reported dwarf galaxy candidates in the DES footprint (red circles; \citealt{Bechtol2015}; \citealt{Drlica2015}; \citealt{Koposov2015}; \citealt{Kim2015b}; \citealt{Luque2017}), other recently reported dwarf galaxy candidates (green diamonds; \citealt{Laevens2015a,Laevens2015b}; \citealt{Martin2015a}; \citealt{Drlica2016}; \citealt{Homma2016,Homma2017}), DES\,3 (red star), and DES\,J0222.7$-$5217 (green star) are shown. The red circles connected with a line represent the two previous DES\,J0222.7$-$5217 estimates. Note that DES\,3 and DES\,J0222.7$-$5217 lie inside the region inhabited by faint star clusters. The uncertainties of both objects are comparable to the symbol size. The dashed lines indicate contours of constant surface brightness at $\mu=\{25,\, 27.5,\, 30\}\,\mathrm{mag\,arcsec^{-2}}$.}\label{fig:magrh}
\end{figure}

In the left panel of Fig. \ref{fig:cmds}, we show the CMD of DES\,3 from the SOAR data. We show only stars within an ellipse with semimajor axis $a\sim 2r_\mathrm{h}$ according to the best-fitting exponential profile (see Table \ref{tab:DES3}). This CMD clearly shows the presence of MS and SGB stars down to $g_\mathrm{DES} \simeq 25.5$. We can also identify one star that may belong to the HB. In the middle panel of Fig. \ref{fig:cmds}, we show the CMD of background stars contained in an elliptical annulus, centred on DES\,3, of equal area as the previous panel, whose inner semimajor axis is equal to $a= 2\arcmin$. We used the technique described in Section \ref{sec:data} to determine the effective area of this region. The excess of stars within the isochrone filter seen in the left panel relative to the background is remarkable, attesting not only the physical reality of DES\,3, but also allowing a detailed CMD analysis. For comparison, in the right panel of this figure, we show the CMD of DES\,3 from the DES data (see right panel of Fig. \ref{fig:mapden}). As expected, the SOAR-based CMD is substantially more informative.

By using the maximum-likelihood method to fit the CMD distribution, we find that DES\,3 population is well described by a \textsc{parsec} \citep{Bressan2012} isochrone model with age $9.8\,\mathrm{Gyr}$, $(m-M)_0\simeq 19.41$, and $[\mathrm{Fe/H}]\simeq -1.88$. These parameters agree for both density profile models (exponential and Plummer). This is not surprising given that these models adequately describe the observed density profile of DES\,3 (see Fig. \ref{fig:profile}). The best-fitting isochrone is overplotted in each panel of Fig. \ref{fig:cmds}  as the solid line. In the same figure, an isochrone filter (gray shaded area) is also shown in each panel. 

The absolute magnitude ($M_V$) has been determined using a similar approach as \citet{Koposov2015}. We integrate over all masses along the best-fitting model isochrone assuming a \citet{Kroupa2001} initial mass function, and normalize the number of objects by those observed in the CMD with $r_\mathrm{DES} < 24.5\,\mathrm{mag}$ and which fall in the isochrone filter. For this estimate, we selected all stars with a membership threshold of $p\geqslant 0.01$ taken from the best profile fits. We corrected the star counts for completeness by weighting each star by $w_i= 1/c_i$, where $c_i$ is the completeness of the star interpolated in magnitude for an interval of radius (see Fig. \ref{fig:comp}). Due to the low number of stars observed in this type of objects, the estimate of the absolute magnitude has large uncertainty. We then calculate the uncertainty by estimating the upper and lower limits for the integrated $V$ magnitude. We convert the $g_\mathrm{DES}$ and $r_\mathrm{DES}$ magnitudes to $V$ magnitude using a SDSS stellar calibration sample\footnote{This new version of transformation equations are based on SDSS Data Release 13 (DR13) and DES Year 3 Annual Release (Y3A1) single-epoch data.} and the equation from \citet{Jester2005},
\begin{align}
\begin{aligned}
g_\mathrm{DES} &= g_\mathrm{SDSS}-0.075(g_\mathrm{SDSS}-r_\mathrm{SDSS})+0.001\\
r_\mathrm{DES} &= r_\mathrm{SDSS}-0.069(g_\mathrm{SDSS}-r_\mathrm{SDSS})-0.009\\
V &= g_\mathrm{SDSS}-0.59(g_\mathrm{SDSS} - r_\mathrm{SDSS})-0.01.\label{eq:2}
\end{aligned}
\end{align}
This procedure yields an absolute magnitude of $M_V= -1.8^{+0.4}_{-0.3}$ for the  exponencial model and $M_V=-2.0^{+0.4}_{-0.3}$ for the Plummer model. Therefore, in the size-luminosity plane DES\,3 lies in the faint star cluster region (see Fig. \ref{fig:magrh}). The luminosity of DES\,3 is comparable to Koposov\,1 ($M_V\sim -2$; \citealt{Koposov2007}), DES\,1 ($M_V\sim -2.21$; \citealt{Luque2016}), and Gaia\,2 ($M_V\simeq -2$; \citealt{Koposov2017}). However, the small size ($r_\mathrm{h}\sim 6.5\,\mathrm{pc}$) of DES\,3 is comparable to  Balbinot\,1 ($r_\mathrm{h} \simeq 7.24\,\mathrm{pc}$; \citealt{Balbinot2013}), Kim\,1 ($r_\mathrm{h} \simeq 6.9\,\mathrm{pc}$; \citealt{Kim2015a}) and Laevens\,3 ($r_\mathrm{h} \simeq 7\,\mathrm{pc}$; \citealt{Laevens2015b}).

\section{Properties of DES\,J0222.7$-$5217}
\label{sec:DES4}
We obtained deeper photometric data for DES\,J0222.7$-$5217 with the SOAR telescope. Much like our SOAR imaging of DES\,3, our SOAR imaging of DES\,J0222.7$-$5217 is $\sim 1\,\mathrm{mag}$ deeper than the DES data. We apply the same methodology described in Section \ref{sec:param} to provide updated properties of DES\,J0222.7$-$5217.

\begin{figure}
  \includegraphics[width=.49\textwidth]{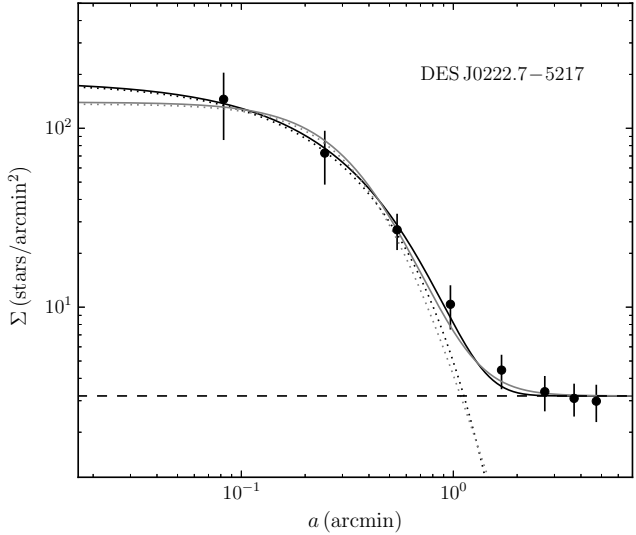}
  \caption{Elliptical surface density profile of DES\,J0222.7$-$5217. The filled points represent the observed values with $1\sigma$ error bars. The gray (black) dotted line represents the best-fitting of Plummer (exponential) profile. The horizontal dashed line represents the background density. The gray (black) solid line is the combination of the background level with the Plummer (exponential) profile.}\label{fig:denDES4}
\end{figure}

Fig. \ref{fig:denDES4} shows the binned elliptical density profile of DES\,J0222.7$-$5217 and the best-fitting exponential and Plummer models. We find that the two models yield very similar structural parameters for this object (see Table \ref{fitpars_DES4}). Our half-light radius ($r_\mathrm{h}\sim 0\farcm 47$) estimate is $\sim 12$ per cent larger than the value determined by \citet[][$r_\mathrm{h}\simeq 0\farcm 42$]{Bechtol2015}, but it is $\sim 13$ per cent smaller than that determined by \citet[][$r_\mathrm{h}\simeq 0\farcm 54$]{Koposov2015}. We find the system to be more elliptical ($\epsilon \sim 0.53$) and less rotated ($\theta\sim -84\,\mathrm{deg}$) when compared to the values previously determined ($\epsilon\simeq 0.27$ and $\theta\simeq -97\,\mathrm{deg}$; \citealt{Koposov2015}). In general, our results are in agreement within $1\sigma$ with the literature. However, our deeper data has allowed us to better constrain the structural parameters of DES\,J0222.7$-$5217 thereby reducing significantly the uncertainties reported in previous work.

\begin{table}
\caption{Properties of DES\,J0222.7$-$5217.}\label{tab:DES4}
\hspace{-.6cm}
\scalebox{.92}{
\begin{tabular}{lccc}\hline
Parameters & Exponential profile & Plummer profile & Unit\\\hline
$\alpha_0\,(J2000)$ & $02\;22\;45.46^{+0.16}_{-0.19}$  & $02\;22\;45.56^{+0.20}_{-0.20}$ & $\mathrm{h\;m\;s}$\\
$\delta_0\,(J2000)$ & $-52\;17\;06.00^{+1.44}_{-1.44}$ & $-52\;17\;06.06^{+1.38}_{-1.38}$  & $\degr\;\arcmin\;\arcsec$\\
$l$ & $274.96$ & $274.96$  &  $\mathrm{deg}$\\
$b$ & $-59.60$ & $-59.60$  & $\mathrm{deg}$\\
$\mathrm{D}_{\sun}$ & $77.6^{+2.1}_{-3.2}$ & $77.6^{+2.1}_{-2.9}$  &  $\mathrm{kpc}$\\
$r_\mathrm{h}$ & $0.47^{+0.06}_{-0.05}$ & $0.46^{+0.06}_{-0.05}$ & $\mathrm{arcmin}$ \\
$r_\mathrm{h}$ & $10.61^{+1.38}_{-1.21}$\textsuperscript{$a$} & $10.38^{+1.38}_{-1.19}$\textsuperscript{$a$}&  $\mathrm{pc}$\\
$\theta$& $-83.2^{+6.7}_{-6.2}$ & $-84.6^{+6.1}_{-5.6}$ &  $\mathrm{deg}$\\
$\epsilon$& $0.52^{+0.07}_{-0.09}$ & $0.53^{+0.07}_{-0.09}$ &  \\
$\Sigma_{\mathrm{bgd}}$& $3.2 \pm 0.4$ & $3.2 \pm 0.4$ & $\mathrm{\frac{stars}{arcmin^2}}$\\
$M_V$ & $-2.6^{+0.5}_{-0.3}$ & $-2.8^{+0.4}_{-0.3}$ & $\mathrm{mag}$ \\
$\mathrm{[Fe/H]}$ & $-2.01^{+0.38}_{-0.12}$ & $-2.01^{+0.23}_{-0.12}$  & $\mathrm{dex}$  \\
$\mathrm{Age}$& $12.6^{+0.6}_{-0.6}$ & $12.6^{+0.3}_{-0.6}$  & $\mathrm{Gyr}$ \\
$(m-M)_0$ & $19.45^{+0.06}_{-0.09}$ & $19.45^{+0.06}_{-0.08}$ & $\mathrm{mag}$\\\hline
\end{tabular}}\label{fitpars_DES4}\\
\leftline{\scalebox{.91}{\textit{Note.} \textsuperscript{$a$}\footnotesize{Adopting a distance of $77.6\,\mathrm{kpc}$.}}}
\end{table}

\begin{figure*}
  \includegraphics[width=1.\textwidth]{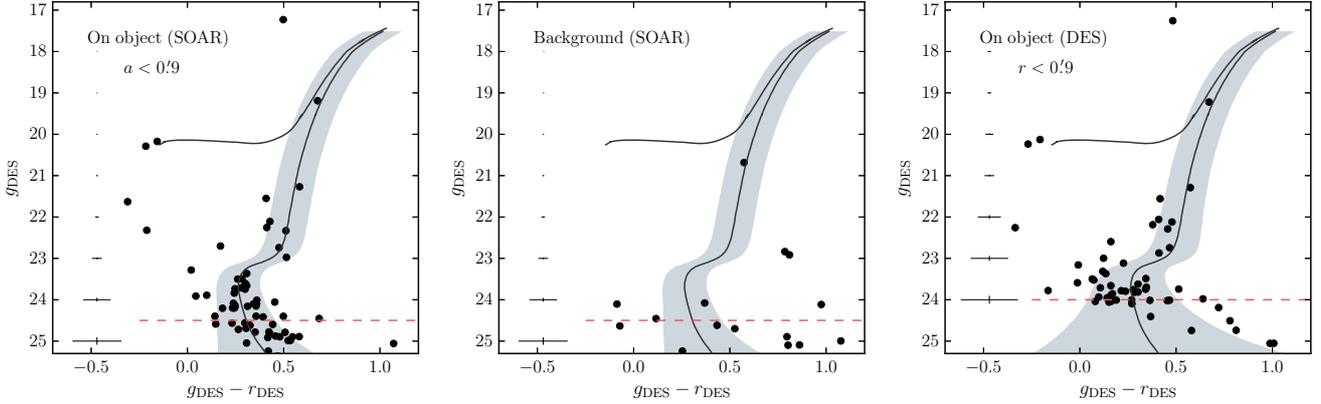}
  \caption{Left panel: CMD of DES\,J0222.7$-$5217 from the SOAR data. Only stars inside an ellipse with semi-major axis $a\sim 2r_\mathrm{h}$ from the centre of DES\,J0222.7$-$5217 are shown. In this and the other two panels, the best-fitting \textsc{parsec} \citep{Bressan2012} isochrone  derived from the SOAR data is shown. The isochrone filter (gray shaded area) based on photometric uncertainties contains the most likely members. Middle panel: CMD of field stars in an elliptical annulus of equal area on the sky as the previous panel. Right panel: CMD of DES stars within a circle with radius $r=0\farcm 9$ from the centre of DES\,J0222.7$-$5217. The horizontal dashed line in each panel indicates the faint magnitude limit used. The mean photometric errors in both colour and magnitude are shown in the extreme left of each panel.}\label{fig:cmdsDES4}
\end{figure*}

The CMD of DES\,J0222.7$-$5217 from the SOAR data is shown in the left panel of Fig. \ref{fig:cmdsDES4}. The CMD, built with stars within an elliptical annulus of semimajor axis $a\sim 2r_\mathrm{h}$ centred on the object, clearly shows MS, MSTO, blue straggler (BS), red giant branch (RGB), and HB stars. Note that there is a potential asymptotic giant branch (AGB) star. The middle panel of Fig. \ref{fig:cmdsDES4} shows the CMD of field stars contained in an elliptical annulus of equal area as the previous panel, whose inner semimajor axis is $a=2\farcm 5$. For comparison, in the right panel of Fig. \ref{fig:cmdsDES4}, we show the CMD of stars within $r=0\farcm 9$ of the centre of DES\,J0222.7$-$5217 from the DES data. Much like SOAR data, this CMD shows BS, RGB, HB, and  AGB stars, however, it does not provide enough information about the MSTO and MS stars. Therefore, the SOAR CMD is substantially more informative than the DES CMD.

The best-fitting model isochrone, determined from the our CMD fit method, estimates that DES\,J0222.7$-$5217 is located at a distance of $\mathrm{D}_{\sun}\simeq 77.6\,\mathrm{kpc}$ and its stellar population is old ($\simeq 12.6\,\mathrm{Gyr}$) and metal-poor ($[\mathrm{Fe/H]}\simeq -2.01$). Again, the values of these parameters are consistent for both density profile models (see Table \ref{tab:DES4}).
 
There is a moderate discrepancy between our heliocentric distance estimate and previous estimates of $\mathrm{D}_{\sun}\simeq 95\,\mathrm{kpc}$ \citep{Bechtol2015} and $\mathrm{D}_{\sun}\simeq 87\,\mathrm{kpc}$ \citep{Koposov2015}. In fact, this discrepancy may be due mainly to the limiting magnitude used in this work and those used by \citet{Bechtol2015} and \citet{Koposov2015}, since the resolution of the MS and MSTO allows for an improved distance measurement and estimations of the age and metallicity. Unfortunately, the metallicity and age values were not reported by these authors.

We estimate an absolute magnitude for DES\,J0222.7$-$5217 of $M_V=-2.6^{+0.5}_{-0.3}$ for the exponential profile and $M_V=-2.8^{+0.4}_{-0.3}$ for the Plummer profile.  For these determinations, we used stars brighter than $r_\mathrm{DES} =24\,\mathrm{mag}$ and the star counts 
were corrected for sample incompleteness. These results are in agreement (within $1\sigma$) with those previously reported in the literature, $M_V\simeq -2.4 $ (\citealt{Bechtol2015}) and $M_V\simeq -2.0$ (\citealt{Koposov2015}). 

With a half-light radius of $r_\mathrm{h}\sim 10.5\,\mathrm{pc}$ and a luminosity of $M_V\sim -2.7$, DES\,J0222.7$-$5217 lies in a region of size-luminosity space occupied by faint star clusters (see Fig. \ref{fig:magrh}). Interestingly, the half-light radius, ellipticity and absolute magnitude of DES\,J0222.7$-$5217 are comparable to those for the star cluster DES\,1 ($r_\mathrm{h}\simeq 9.88\,\mathrm{pc}$, $\epsilon\simeq 0.53$ and $M_V\simeq -2.21$; \citealt{Luque2016}).

\section{Conclusions}
\label{sec:conclusions}
In this paper, we announce the discovery of a new MW faint star cluster found in DES\, Y1A1 data, which we name DES\,3. Its confirmation as a physical system required deep photometric imaging from the SOAR telescope. This new object adds to the 21 systems that have been found in the first two years of DES \citep{Bechtol2015,Drlica2015,Koposov2015,Kim2015b,Luque2016,Luque2017}. 

With a MCMC technique and two different density profile models (exponential and Plummer), we find that DES\,3 is compact ($r_\mathrm{h}\sim 0\farcm 3$) and slightly elongated ($\epsilon\sim 0.15$). The morphology of DES\,3 does not suggest any evidence of on-going tidal disruption. 

By means of an isochrone fit, we derive a distance of $\simeq 76.2\,\mathrm{kpc}$ for DES\,3. It is consistent with being dominated by an old ($\simeq 9.8\,\mathrm{Gyr}$) and metal-poor ($\mathrm{[Fe/H]}\simeq -1.88$) population, as commonly observed in MW GCs found in the Galactic halo. However, its small physical size ($r_\mathrm{h}\sim 6.5\,\mathrm{pc}$) and low luminosity ($M_V\sim -1.9$) place DES\,3 in the region occupied by faint star clusters as observed in Fig. \ref{fig:magrh}. In fact, DES\,3 is also one of the faint star clusters with smallest angular size known so far.

With deep SOAR data we found that DES\,J0222.7$-$5217 is located at a heliocentric distance of $77.6\,\mathrm{kpc} $, and it hosts an old ($\simeq 12.6\,\mathrm{Gyr}$) and metal-poor ($\mathrm{[Fe/H]}\simeq -2.01$) stellar population. Our best-fitting structural parameters for DES\,J0222.7$-$5217 are in general agreement (within $1\sigma$) with the ones derived by \citet{Bechtol2015} and \citet{Koposov2015}, although the heliocentric distance determined in this work points to a closer object than previously reported.
The half-light radius ($r_\mathrm{h}\sim 10.5\,\mathrm{pc}$) and luminosity ($M_V \sim -2.7$) of DES\,J0222.7$-$5217 suggest that it could be classified as a faint star cluster. However, the spectroscopic determination of the radial velocity of DES\,J0222.7$-$5217 will be very useful to confirm its nature.
 
Based on the Magellanic Stream \citep[MS;][]{Nidever2008} coordinates of DES\,3, $(L_\mathrm{MS},\,B_\mathrm{MS})=-37\fdg 39,\,-31\fdg 69$ and DES\,J0222.7$-$5217, $(L_\mathrm{MS},\,B_\mathrm{MS})=-26\fdg 45,\, 8\fdg 25$, it is interesting to note that DES\,J0222.7$-$5217 is in a region where there is a high probability of finding objects associated with the Magellanic Clouds, while DES\,3 lies close to a sequence of faint dwarf galaxies, some of which may also be associated with the Clouds (Fig. 10; \citealt{Jethwa2016}). However, DES\,3 lies outside $\pm 20\degr$ of the plane of the Magellanic Stream where the satellites of the LMC would be distributed \citep{Jethwa2016}.

Finally, the discovery of DES\,3 in DES data indicates that the census of stellar systems, with characteristics of faint star clusters, is still incomplete. It is likely that additional new stellar systems will be found in future DES data.  We have demonstrated the value of deeper imaging to improve the photometric errors and to detect stars at and below the MSTO. This greatly improves the fitting of isochrones, in particular tightening the constraints on the age. For many of these newly discovered objects a wide field is not needed and SOI on the SOAR telescope is an ideal instrument for follow-up studies.

\section*{Acknowledgements} 
This paper has gone through internal review by the DES collaboration.

Funding for the DES Projects has been provided by the U.S. Department of Energy, the U.S. National Science Foundation, the Ministry of Science and Education of Spain, 
the Science and Technology Facilities Council of the United Kingdom, the Higher Education Funding Council for England, the National Center for Supercomputing 
Applications at the University of Illinois at Urbana-Champaign, the Kavli Institute of Cosmological Physics at the University of Chicago, 
the Center for Cosmology and Astro-Particle Physics at the Ohio State University,
the Mitchell Institute for Fundamental Physics and Astronomy at Texas A\&M University, Financiadora de Estudos e Projetos, 
Funda{\c c}{\~a}o Carlos Chagas Filho de Amparo {\`a} Pesquisa do Estado do Rio de Janeiro, Conselho Nacional de Desenvolvimento Cient{\'i}fico e Tecnol{\'o}gico and 
the Minist{\'e}rio da Ci{\^e}ncia, Tecnologia e Inova{\c c}{\~a}o, the Deutsche Forschungsgemeinschaft and the Collaborating Institutions in the Dark Energy Survey. 

The Collaborating Institutions are Argonne National Laboratory, the University of California at Santa Cruz, the University of Cambridge, Centro de Investigaciones Energ{\'e}ticas, 
Medioambientales y Tecnol{\'o}gicas-Madrid, the University of Chicago, University College London, the DES-Brazil Consortium, the University of Edinburgh, 
the Eidgen{\"o}ssische Technische Hochschule (ETH) Z{\"u}rich, 
Fermi National Accelerator Laboratory, the University of Illinois at Urbana-Champaign, the Institut de Ci{\`e}ncies de l'Espai (IEEC/CSIC), 
the Institut de F{\'i}sica d'Altes Energies, Lawrence Berkeley National Laboratory, the Ludwig-Maximilians Universit{\"a}t M{\"u}nchen and the associated Excellence Cluster Universe, 
the University of Michigan, the National Optical Astronomy Observatory, the University of Nottingham, The Ohio State University, the University of Pennsylvania, the University of Portsmouth, 
SLAC National Accelerator Laboratory, Stanford University, the University of Sussex, Texas A\&M University, and the OzDES Membership Consortium.

Based in part on observations at Cerro Tololo Inter-American Observatory, National Optical Astronomy Observatory, which is operated by the Association of 
Universities for Research in Astronomy (AURA) under a cooperative agreement with the National Science Foundation.

The DES data management system is supported by the National Science Foundation under Grant Numbers AST-1138766 and AST-1536171.
The DES participants from Spanish institutions are partially supported by MINECO under grants AYA2015-71825, ESP2015-88861, FPA2015-68048, SEV-2012-0234, SEV-2016-0597, and MDM-2015-0509, 
some of which include ERDF funds from the European Union. IFAE is partially funded by the CERCA program of the Generalitat de Catalunya.
Research leading to these results has received funding from the European Research
Council under the European Union's Seventh Framework Program (FP7/2007-2013) including ERC grant agreements 240672, 291329, and 306478.
We  acknowledge support from the Australian Research Council Centre of Excellence for All-sky Astrophysics (CAASTRO), through project number CE110001020.

This manuscript has been authored by Fermi Research Alliance, LLC under Contract No. DE-AC02-07CH11359 with the U.S. Department of Energy, Office of Science, Office of High Energy Physics. The United States Government retains and the publisher, by accepting the article for publication, acknowledges that the United States Government retains a non-exclusive, paid-up, irrevocable, world-wide license to publish or reproduce the published form of this manuscript, or allow others to do so, for United States Government purposes.

Based on observations obtained at the Southern Astrophysical Research (SOAR) telescope, which is a joint project of the Minist\'{e}rio da Ci\^{e}ncia, Tecnologia, e Inova\c{c}\~{a}o (MCTI) da Rep\'{u}blica Federativa do Brasil, the U.S. National Optical Astronomy Observatory (NOAO), the University of North Carolina at Chapel Hill (UNC), and Michigan State University (MSU).


\bibliographystyle{mnras}          
\bibliography{bib}\vspace*{3em}

\begin{thebibliography}{}
\makeatletter
\relax
\def\mn@urlcharsother{\let\do\@makeother \do\$\do\&\do\#\do\^\do\_\do\%\do\~}
\def\mn@doi{\begingroup\mn@urlcharsother \@ifnextchar [ {\mn@doi@}
  {\mn@doi@[]}}
\def\mn@doi@[#1]#2{\def\@tempa{#1}\ifx\@tempa\@empty \href
  {http://dx.doi.org/#2} {doi:#2}\else \href {http://dx.doi.org/#2} {#1}\fi
  \endgroup}
\def\mn@eprint#1#2{\mn@eprint@#1:#2::\@nil}
\def\mn@eprint@arXiv#1{\href {http://arxiv.org/abs/#1} {{\tt arXiv:#1}}}
\def\mn@eprint@dblp#1{\href {http://dblp.uni-trier.de/rec/bibtex/#1.xml}
  {dblp:#1}}
\def\mn@eprint@#1:#2:#3:#4\@nil{\def\@tempa {#1}\def\@tempb {#2}\def\@tempc
  {#3}\ifx \@tempc \@empty \let \@tempc \@tempb \let \@tempb \@tempa \fi \ifx
  \@tempb \@empty \def\@tempb {arXiv}\fi \@ifundefined
  {mn@eprint@\@tempb}{\@tempb:\@tempc}{\expandafter \expandafter \csname
  mn@eprint@\@tempb\endcsname \expandafter{\@tempc}}}

\bibitem[\protect\citeauthoryear{{Balbinot} et~al.,}{{Balbinot}
  et~al.}{2013}]{Balbinot2013}
{Balbinot} E.,  et~al., 2013, \mn@doi [\apj] {10.1088/0004-637X/767/2/101},
  \href {http://adsabs.harvard.edu/abs/2013ApJ...767..101B} {767, 101}

\bibitem[\protect\citeauthoryear{{Balbinot} et~al.,}{{Balbinot}
  et~al.}{2015}]{Balbinot2015}
{Balbinot} E.,  et~al., 2015, \mn@doi [\mnras] {10.1093/mnras/stv356}, \href
  {http://adsabs.harvard.edu/abs/2015MNRAS.449.1129B} {449, 1129}

\bibitem[\protect\citeauthoryear{{Bechtol} et~al.,}{{Bechtol}
  et~al.}{2015}]{Bechtol2015}
{Bechtol} K.,  et~al., 2015, \mn@doi [\apj] {10.1088/0004-637X/807/1/50}, \href
  {http://adsabs.harvard.edu/abs/2015ApJ...807...50B} {807, 50}

\bibitem[\protect\citeauthoryear{{Belokurov} et~al.,}{{Belokurov}
  et~al.}{2010}]{Belokurov2010}
{Belokurov} V.,  et~al., 2010, \mn@doi [\apjl] {10.1088/2041-8205/712/1/L103},
  \href {http://adsabs.harvard.edu/abs/2010ApJ...712L.103B} {712, L103}

\bibitem[\protect\citeauthoryear{{Bertin}}{{Bertin}}{2011}]{Bertin2011}
{Bertin} E.,  2011, in {Evans} I.~N.,  {Accomazzi} A.,  {Mink} D.~J.,   {Rots}
  A.~H.,  eds,  Astronomical Society of the Pacific Conference Series Vol. 442,
  Astronomical Data Analysis Software and Systems XX. p.~435

\bibitem[\protect\citeauthoryear{{Bertin} \& {Arnouts}}{{Bertin} \&
  {Arnouts}}{1996}]{Bertin1996}
{Bertin} E.,  {Arnouts} S.,  1996, \mn@doi [\aaps] {10.1051/aas:1996164}, \href
  {http://adsabs.harvard.edu/abs/1996A%26AS..117..393B} {117, 393}

\bibitem[\protect\citeauthoryear{{Bressan}, {Marigo}, {Girardi}, {Salasnich},
  {Dal Cero}, {Rubele}  \& {Nanni}}{{Bressan} et~al.}{2012}]{Bressan2012}
{Bressan} A.,  {Marigo} P.,  {Girardi} L.,  {Salasnich} B.,  {Dal Cero} C.,
  {Rubele} S.,   {Nanni} A.,  2012, \mn@doi [\mnras]
  {10.1111/j.1365-2966.2012.21948.x}, \href
  {http://adsabs.harvard.edu/abs/2012MNRAS.427..127B} {427, 127}

\bibitem[\protect\citeauthoryear{{Carraro} \& {Bensby}}{{Carraro} \&
  {Bensby}}{2009}]{Carraro2009}
{Carraro} G.,  {Bensby} T.,  2009, \mn@doi [\mnras]
  {10.1111/j.1745-3933.2009.00694.x}, \href
  {http://adsabs.harvard.edu/abs/2009MNRAS.397L.106C} {397, L106}

\bibitem[\protect\citeauthoryear{{Davis}, {Efstathiou}, {Frenk}  \&
  {White}}{{Davis} et~al.}{1985}]{Davis1985}
{Davis} M.,  {Efstathiou} G.,  {Frenk} C.~S.,   {White} S.~D.~M.,  1985,
  \mn@doi [\apj] {10.1086/163168}, \href
  {http://adsabs.harvard.edu/abs/1985ApJ...292..371D} {292, 371}

\bibitem[\protect\citeauthoryear{{Desai} et~al.,}{{Desai}
  et~al.}{2012}]{Desai2012}
{Desai} S.,  et~al., 2012, \mn@doi [\apj] {10.1088/0004-637X/757/1/83}, \href
  {http://adsabs.harvard.edu/abs/2012ApJ...757...83D} {757, 83}

\bibitem[\protect\citeauthoryear{{Dooley}, {Peter}, {Carlin}, {Frebel},
  {Bechtol}  \& {Willman}}{{Dooley} et~al.}{2017}]{Dooley2017}
{Dooley} G.~A.,  {Peter} A.~H.~G.,  {Carlin} J.~L.,  {Frebel} A.,  {Bechtol}
  K.,   {Willman} B.,  2017, preprint, \href
  {http://adsabs.harvard.edu/abs/2017arXiv170305321D} {} (\mn@eprint {arXiv}
  {1703.05321})

\bibitem[\protect\citeauthoryear{{Drlica-Wagner} et~al.,}{{Drlica-Wagner}
  et~al.}{2015}]{Drlica2015}
{Drlica-Wagner} A.,  et~al., 2015, \mn@doi [\apj]
  {10.1088/0004-637X/813/2/109}, \href
  {http://adsabs.harvard.edu/abs/2015ApJ...813..109D} {813, 109}

\bibitem[\protect\citeauthoryear{{Drlica-Wagner} et~al.,}{{Drlica-Wagner}
  et~al.}{2016}]{Drlica2016}
{Drlica-Wagner} A.,  et~al., 2016, \mn@doi [\apjl]
  {10.3847/2041-8205/833/1/L5}, \href
  {http://adsabs.harvard.edu/abs/2016ApJ...833L...5D} {833, L5}

\bibitem[\protect\citeauthoryear{{Drlica-Wagner} et~al.,}{{Drlica-Wagner}
  et~al.}{2017}]{Drlica-Wagner2017}
{Drlica-Wagner} A.,  et~al., 2017, preprint, \href
  {http://adsabs.harvard.edu/abs/2017arXiv170801531D} {} (\mn@eprint {arXiv}
  {1708.01531})

\bibitem[\protect\citeauthoryear{{Fadely}, {Willman}, {Geha}, {Walsh},
  {Mu{\~n}oz}, {Jerjen}, {Vargas}  \& {Da Costa}}{{Fadely}
  et~al.}{2011}]{Fadely2011}
{Fadely} R.,  {Willman} B.,  {Geha} M.,  {Walsh} S.,  {Mu{\~n}oz} R.~R.,
  {Jerjen} H.,  {Vargas} L.~C.,   {Da Costa} G.~S.,  2011, \mn@doi [\aj]
  {10.1088/0004-6256/142/3/88}, \href
  {http://adsabs.harvard.edu/abs/2011AJ....142...88F} {142, 88}

\bibitem[\protect\citeauthoryear{{Flaugher} et~al.,}{{Flaugher}
  et~al.}{2015}]{Flaugher2015}
{Flaugher} B.,  et~al., 2015, \mn@doi [\aj] {10.1088/0004-6256/150/5/150},
  \href {http://adsabs.harvard.edu/abs/2015AJ....150..150F} {150, 150}

\bibitem[\protect\citeauthoryear{{Font} et~al.,}{{Font}
  et~al.}{2011}]{Font2011}
{Font} A.~S.,  et~al., 2011, \mn@doi [\mnras]
  {10.1111/j.1365-2966.2011.19339.x}, \href
  {http://adsabs.harvard.edu/abs/2011MNRAS.417.1260F} {417, 1260}

\bibitem[\protect\citeauthoryear{{Foreman-Mackey}, {Hogg}, {Lang}  \&
  {Goodman}}{{Foreman-Mackey} et~al.}{2013}]{Foreman2013}
{Foreman-Mackey} D.,  {Hogg} D.~W.,  {Lang} D.,   {Goodman} J.,  2013, \mn@doi
  [\pasp] {10.1086/670067}, \href
  {http://adsabs.harvard.edu/abs/2013PASP..125..306F} {125, 306}

\bibitem[\protect\citeauthoryear{{Harris}}{{Harris}}{2010}]{Harris2010}
{Harris} W.~E.,  2010, preprint, \href
  {http://adsabs.harvard.edu/abs/2010arXiv1012.3224H} {} (\mn@eprint {arXiv}
  {1012.3224})

\bibitem[\protect\citeauthoryear{{Homma} et~al.,}{{Homma}
  et~al.}{2016}]{Homma2016}
{Homma} D.,  et~al., 2016, \mn@doi [\apj] {10.3847/0004-637X/832/1/21}, \href
  {http://adsabs.harvard.edu/abs/2016ApJ...832...21H} {832, 21}

\bibitem[\protect\citeauthoryear{{Homma} et~al.,}{{Homma}
  et~al.}{2017}]{Homma2017}
{Homma} D.,  et~al., 2017, preprint, \href
  {http://adsabs.harvard.edu/abs/2017arXiv170405977H} {} (\mn@eprint {arXiv}
  {1704.05977})

\bibitem[\protect\citeauthoryear{{Ibata}, {Gilmore}  \& {Irwin}}{{Ibata}
  et~al.}{1994}]{Ibata1994}
{Ibata} R.~A.,  {Gilmore} G.,   {Irwin} M.~J.,  1994, \mn@doi [\nat]
  {10.1038/370194a0}, \href {http://adsabs.harvard.edu/abs/1994Natur.370..194I}
  {370, 194}

\bibitem[\protect\citeauthoryear{{Ibata}, {Nipoti}, {Sollima}, {Bellazzini},
  {Chapman}  \& {Dalessandro}}{{Ibata} et~al.}{2013}]{Ibata2013}
{Ibata} R.,  {Nipoti} C.,  {Sollima} A.,  {Bellazzini} M.,  {Chapman} S.~C.,
  {Dalessandro} E.,  2013, \mn@doi [\mnras] {10.1093/mnras/sts302}, \href
  {http://adsabs.harvard.edu/abs/2013MNRAS.428.3648I} {428, 3648}

\bibitem[\protect\citeauthoryear{{Jester} et~al.,}{{Jester}
  et~al.}{2005}]{Jester2005}
{Jester} S.,  et~al., 2005, \mn@doi [\aj] {10.1086/432466}, \href
  {http://adsabs.harvard.edu/abs/2005AJ....130..873J} {130, 873}

\bibitem[\protect\citeauthoryear{{Jethwa}, {Erkal}  \& {Belokurov}}{{Jethwa}
  et~al.}{2016}]{Jethwa2016}
{Jethwa} P.,  {Erkal} D.,   {Belokurov} V.,  2016, \mn@doi [\mnras]
  {10.1093/mnras/stw1343}, \href
  {http://adsabs.harvard.edu/abs/2016MNRAS.461.2212J} {461, 2212}

\bibitem[\protect\citeauthoryear{{Kim} \& {Jerjen}}{{Kim} \&
  {Jerjen}}{2015a}]{Kim2015a}
{Kim} D.,  {Jerjen} H.,  2015a, \mn@doi [\apj] {10.1088/0004-637X/799/1/73},
  \href {http://adsabs.harvard.edu/abs/2015ApJ...799...73K} {799, 73}

\bibitem[\protect\citeauthoryear{{Kim} \& {Jerjen}}{{Kim} \&
  {Jerjen}}{2015b}]{Kim2015b}
{Kim} D.,  {Jerjen} H.,  2015b, \mn@doi [\apjl] {10.1088/2041-8205/808/2/L39},
  \href {http://adsabs.harvard.edu/abs/2015ApJ...808L..39K} {808, L39}

\bibitem[\protect\citeauthoryear{{Kim}, {Jerjen}, {Milone}, {Mackey}  \& {Da
  Costa}}{{Kim} et~al.}{2015}]{Kim22015}
{Kim} D.,  {Jerjen} H.,  {Milone} A.~P.,  {Mackey} D.,   {Da Costa} G.~S.,
  2015, \mn@doi [\apj] {10.1088/0004-637X/803/2/63}, \href
  {http://adsabs.harvard.edu/abs/2015ApJ...803...63K} {803, 63}

\bibitem[\protect\citeauthoryear{{Kim}, {Jerjen}, {Mackey}, {Da Costa}  \&
  {Milone}}{{Kim} et~al.}{2016a}]{Kim32016}
{Kim} D.,  {Jerjen} H.,  {Mackey} D.,  {Da Costa} G.~S.,   {Milone} A.~P.,
  2016a, \mn@doi [\apj] {10.3847/0004-637X/820/2/119}, \href
  {http://adsabs.harvard.edu/abs/2016ApJ...820..119K} {820, 119}

\bibitem[\protect\citeauthoryear{{Kim} et~al.,}{{Kim} et~al.}{2016b}]{Kim2016b}
{Kim} D.,  et~al., 2016b, \mn@doi [\apj] {10.3847/0004-637X/833/1/16}, \href
  {http://adsabs.harvard.edu/abs/2016ApJ...833...16K} {833, 16}

\bibitem[\protect\citeauthoryear{{Koposov} et~al.,}{{Koposov}
  et~al.}{2007}]{Koposov2007}
{Koposov} S.,  et~al., 2007, \mn@doi [\apj] {10.1086/521422}, \href
  {http://adsabs.harvard.edu/abs/2007ApJ...669..337K} {669, 337}

\bibitem[\protect\citeauthoryear{{Koposov}, {Belokurov}, {Torrealba}  \&
  {Evans}}{{Koposov} et~al.}{2015a}]{Koposov2015}
{Koposov} S.~E.,  {Belokurov} V.,  {Torrealba} G.,   {Evans} N.~W.,  2015a,
  \mn@doi [\apj] {10.1088/0004-637X/805/2/130}, \href
  {http://adsabs.harvard.edu/abs/2015ApJ...805..130K} {805, 130}

\bibitem[\protect\citeauthoryear{{Koposov} et~al.,}{{Koposov}
  et~al.}{2015b}]{Koposov2015b}
{Koposov} S.~E.,  et~al., 2015b, \mn@doi [\apj] {10.1088/0004-637X/811/1/62},
  \href {http://adsabs.harvard.edu/abs/2015ApJ...811...62K} {811, 62}

\bibitem[\protect\citeauthoryear{{Koposov}, {Belokurov}  \&
  {Torrealba}}{{Koposov} et~al.}{2017}]{Koposov2017}
{Koposov} S.~E.,  {Belokurov} V.,   {Torrealba} G.,  2017, \mn@doi [\mnras]
  {10.1093/mnras/stx1182}, \href
  {http://adsabs.harvard.edu/abs/2017MNRAS.470.2702K} {470, 2702}

\bibitem[\protect\citeauthoryear{{Kroupa}}{{Kroupa}}{2001}]{Kroupa2001}
{Kroupa} P.,  2001, \mn@doi [\mnras] {10.1046/j.1365-8711.2001.04022.x}, \href
  {http://adsabs.harvard.edu/abs/2001MNRAS.322..231K} {322, 231}

\bibitem[\protect\citeauthoryear{{Laevens} et~al.,}{{Laevens}
  et~al.}{2014}]{Laevens2014}
{Laevens} B.~P.~M.,  et~al., 2014, \mn@doi [\apjl]
  {10.1088/2041-8205/786/1/L3}, \href
  {http://adsabs.harvard.edu/abs/2014ApJ...786L...3L} {786, L3}

\bibitem[\protect\citeauthoryear{{Laevens} et~al.,}{{Laevens}
  et~al.}{2015a}]{Laevens2015a}
{Laevens} B.~P.~M.,  et~al., 2015a, \mn@doi [\apjl]
  {10.1088/2041-8205/802/2/L18}, \href
  {http://adsabs.harvard.edu/abs/2015ApJ...802L..18L} {802, L18}

\bibitem[\protect\citeauthoryear{{Laevens} et~al.,}{{Laevens}
  et~al.}{2015b}]{Laevens2015b}
{Laevens} B.~P.~M.,  et~al., 2015b, \mn@doi [\apj]
  {10.1088/0004-637X/813/1/44}, \href
  {http://adsabs.harvard.edu/abs/2015ApJ...813...44L} {813, 44}

\bibitem[\protect\citeauthoryear{{Law} \& {Majewski}}{{Law} \&
  {Majewski}}{2010}]{Law2010}
{Law} D.~R.,  {Majewski} S.~R.,  2010, \mn@doi [\apj]
  {10.1088/0004-637X/714/1/229}, \href
  {http://adsabs.harvard.edu/abs/2010ApJ...714..229L} {714, 229}

\bibitem[\protect\citeauthoryear{{Li} et~al.,}{{Li} et~al.}{2017}]{Li2017}
{Li} T.~S.,  et~al., 2017, \mn@doi [\apj] {10.3847/1538-4357/aa6113}, \href
  {http://adsabs.harvard.edu/abs/2017ApJ...838....8L} {838, 8}

\bibitem[\protect\citeauthoryear{{Luque} et~al.,}{{Luque}
  et~al.}{2016}]{Luque2016}
{Luque} E.,  et~al., 2016, \mn@doi [\mnras] {10.1093/mnras/stw302}, \href
  {http://adsabs.harvard.edu/abs/2016MNRAS.458..603L} {458, 603}

\bibitem[\protect\citeauthoryear{{Luque} et~al.,}{{Luque}
  et~al.}{2017}]{Luque2017}
{Luque} E.,  et~al., 2017, \mn@doi [\mnras] {10.1093/mnras/stx405}, \href
  {http://adsabs.harvard.edu/abs/2017MNRAS.468...97L} {468, 97}

\bibitem[\protect\citeauthoryear{{Mackey} \& {Gilmore}}{{Mackey} \&
  {Gilmore}}{2004}]{Mackey2004}
{Mackey} A.~D.,  {Gilmore} G.~F.,  2004, \mn@doi [\mnras]
  {10.1111/j.1365-2966.2004.08343.x}, \href
  {http://adsabs.harvard.edu/abs/2004MNRAS.355..504M} {355, 504}

\bibitem[\protect\citeauthoryear{{Mackey} \& {van den Bergh}}{{Mackey} \& {van
  den Bergh}}{2005}]{Mackey2005}
{Mackey} A.~D.,  {van den Bergh} S.,  2005, \mn@doi [\mnras]
  {10.1111/j.1365-2966.2005.09080.x}, \href
  {http://adsabs.harvard.edu/abs/2005MNRAS.360..631M} {360, 631}

\bibitem[\protect\citeauthoryear{{Marino}, {Milone}  \& et al.}{{Marino}
  et~al.}{2014}]{Marino2014}
{Marino} A.~F.,  {Milone} A.~P.,   et al. 2014, \mn@doi [\mnras]
  {10.1093/mnras/stu1099}, \href
  {http://adsabs.harvard.edu/abs/2014MNRAS.442.3044M} {442, 3044}

\bibitem[\protect\citeauthoryear{{Marino} et~al.,}{{Marino}
  et~al.}{2015}]{Marino2015}
{Marino} A.~F.,  et~al., 2015, \mn@doi [\mnras] {10.1093/mnras/stv420}, \href
  {http://adsabs.harvard.edu/abs/2015MNRAS.450..815M} {450, 815}

\bibitem[\protect\citeauthoryear{{Martin}, {de Jong}  \& {Rix}}{{Martin}
  et~al.}{2008}]{Martin2008}
{Martin} N.~F.,  {de Jong} J.~T.~A.,   {Rix} H.-W.,  2008, \mn@doi [\apj]
  {10.1086/590336}, \href {http://adsabs.harvard.edu/abs/2008ApJ...684.1075M}
  {684, 1075}

\bibitem[\protect\citeauthoryear{{Martin} et~al.,}{{Martin}
  et~al.}{2015}]{Martin2015a}
{Martin} N.~F.,  et~al., 2015, \mn@doi [\apjl] {10.1088/2041-8205/804/1/L5},
  \href {http://adsabs.harvard.edu/abs/2015ApJ...804L...5M} {804, L5}

\bibitem[\protect\citeauthoryear{{McConnachie}}{{McConnachie}}{2012}]{McConnachie2012}
{McConnachie} A.~W.,  2012, \mn@doi [\aj] {10.1088/0004-6256/144/1/4}, \href
  {http://adsabs.harvard.edu/abs/2012AJ....144....4M} {144, 4}

\bibitem[\protect\citeauthoryear{{Milone} et~al.,}{{Milone}
  et~al.}{2014}]{Milone2014}
{Milone} A.~P.,  et~al., 2014, \mn@doi [\apj] {10.1088/0004-637X/785/1/21},
  \href {http://adsabs.harvard.edu/abs/2014ApJ...785...21M} {785, 21}

\bibitem[\protect\citeauthoryear{{Mohr}, {Armstrong}, {Bertin}  \& et
  al.}{{Mohr} et~al.}{2012}]{Mohr2012}
{Mohr} J.~J.,  {Armstrong} R.,  {Bertin} E.,   et al. 2012, in Society of
  Photo-Optical Instrumentation Engineers (SPIE) Conference Series. p.~0
  (\mn@eprint {arXiv} {1207.3189}), \mn@doi{10.1117/12.926785}

\bibitem[\protect\citeauthoryear{{Mu{\~n}oz}, {Geha}, {C{\^o}t{\'e}}, {Vargas},
  {Santana}, {Stetson}, {Simon}  \& {Djorgovski}}{{Mu{\~n}oz}
  et~al.}{2012}]{Munoz2012}
{Mu{\~n}oz} R.~R.,  {Geha} M.,  {C{\^o}t{\'e}} P.,  {Vargas} L.~C.,  {Santana}
  F.~A.,  {Stetson} P.,  {Simon} J.~D.,   {Djorgovski} S.~G.,  2012, \mn@doi
  [\apjl] {10.1088/2041-8205/753/1/L15}, \href
  {http://adsabs.harvard.edu/abs/2012ApJ...753L..15M} {753, L15}

\bibitem[\protect\citeauthoryear{{Nidever}, {Majewski}  \& {Butler
  Burton}}{{Nidever} et~al.}{2008}]{Nidever2008}
{Nidever} D.~L.,  {Majewski} S.~R.,   {Butler Burton} W.,  2008, \mn@doi [\apj]
  {10.1086/587042}, \href {http://adsabs.harvard.edu/abs/2008ApJ...679..432N}
  {679, 432}

\bibitem[\protect\citeauthoryear{{Odenkirchen}, {Grebel}, {Dehnen}, {Rix},
  {Wolf}  \& {Rockosi}}{{Odenkirchen} et~al.}{2002}]{Odenkirchen2002}
{Odenkirchen} M.,  {Grebel} E.~K.,  {Dehnen} W.,  {Rix} H.-W.,  {Wolf} C.,
  {Rockosi} C.~M.,  2002, in {Grebel} E.~K.,  {Brandner} W.,  eds,
  Astronomical Society of the Pacific Conference Series Vol. 285, Modes of Star
  Formation and the Origin of Field Populations. p.~184

\bibitem[\protect\citeauthoryear{{Pieres} et~al.,}{{Pieres}
  et~al.}{2016}]{Pieres2016}
{Pieres} A.,  et~al., 2016, \mn@doi [\mnras] {10.1093/mnras/stw1260}, \href
  {http://adsabs.harvard.edu/abs/2016MNRAS.461..519P} {461, 519}

\bibitem[\protect\citeauthoryear{{Plummer}}{{Plummer}}{1911}]{Plummer1911}
{Plummer} H.~C.,  1911, \mn@doi [\mnras] {10.1093/mnras/71.5.460}, \href
  {http://adsabs.harvard.edu/abs/1911MNRAS..71..460P} {71, 460}

\bibitem[\protect\citeauthoryear{{Rockosi} et~al.,}{{Rockosi}
  et~al.}{2002}]{Rockosi2002}
{Rockosi} C.~M.,  et~al., 2002, \mn@doi [\aj] {10.1086/340957}, \href
  {http://adsabs.harvard.edu/abs/2002AJ....124..349R} {124, 349}

\bibitem[\protect\citeauthoryear{{Sales}, {Navarro}, {Kallivayalil}  \&
  {Frenk}}{{Sales} et~al.}{2017}]{Sales2017}
{Sales} L.~V.,  {Navarro} J.~F.,  {Kallivayalil} N.,   {Frenk} C.~S.,  2017,
  \mn@doi [\mnras] {10.1093/mnras/stw2816}, \href
  {http://adsabs.harvard.edu/abs/2017MNRAS.465.1879S} {465, 1879}

\bibitem[\protect\citeauthoryear{{Schlegel}, {Finkbeiner}  \&
  {Davis}}{{Schlegel} et~al.}{1998}]{Schlegel1998}
{Schlegel} D.~J.,  {Finkbeiner} D.~P.,   {Davis} M.,  1998, \mn@doi [\apj]
  {10.1086/305772}, \href {http://adsabs.harvard.edu/abs/1998ApJ...500..525S}
  {500, 525}

\bibitem[\protect\citeauthoryear{{Sevilla}, {Armstrong}  \& et al.}{{Sevilla}
  et~al.}{2011}]{Sevilla2011}
{Sevilla} I.,  {Armstrong} R.,   et al. 2011, preprint, \href
  {http://adsabs.harvard.edu/abs/2011arXiv1109.6741S} {} (\mn@eprint {arXiv}
  {1109.6741})

\bibitem[\protect\citeauthoryear{{Simon} et~al.,}{{Simon}
  et~al.}{2015}]{Simon2015}
{Simon} J.~D.,  et~al., 2015, \mn@doi [\apj] {10.1088/0004-637X/808/1/95},
  \href {http://adsas.harvard.edu/abs/2015ApJ...808...95S} {808, 95}

\bibitem[\protect\citeauthoryear{{Simon} et~al.,}{{Simon}
  et~al.}{2017}]{Simon2017}
{Simon} J.~D.,  et~al., 2017, \mn@doi [\apj] {10.3847/1538-4357/aa5be7}, \href
  {http://adsabs.harvard.edu/abs/2017ApJ...838...11S} {838, 11}

\bibitem[\protect\citeauthoryear{{Szabo}, {Pierpaoli}, {Dong}, {Pipino}  \&
  {Gunn}}{{Szabo} et~al.}{2011}]{Szabo2011}
{Szabo} T.,  {Pierpaoli} E.,  {Dong} F.,  {Pipino} A.,   {Gunn} J.,  2011,
  \mn@doi [\apj] {10.1088/0004-637X/736/1/21}, \href
  {http://adsabs.harvard.edu/abs/2011ApJ...736...21S} {736, 21}

\bibitem[\protect\citeauthoryear{{The Dark Energy Survey Collaboration}}{{The
  Dark Energy Survey Collaboration}}{2005}]{DES2005}
{The Dark Energy Survey Collaboration} 2005, preprint, \href
  {http://adsabs.harvard.edu/abs/2005astro.ph.10346T} {} (\mn@eprint {}
  {astro-ph/0510346})

\bibitem[\protect\citeauthoryear{{Torrealba} et~al.,}{{Torrealba}
  et~al.}{2016a}]{Torrealba2016a}
{Torrealba} G.,  et~al., 2016a, \mn@doi [\mnras] {10.1093/mnras/stw2051}, \href
  {http://adsabs.harvard.edu/abs/2016MNRAS.tmp.1171T} {}

\bibitem[\protect\citeauthoryear{{Torrealba}, {Koposov}, {Belokurov}  \&
  {Irwin}}{{Torrealba} et~al.}{2016b}]{Torrealba2016b}
{Torrealba} G.,  {Koposov} S.~E.,  {Belokurov} V.,   {Irwin} M.,  2016b,
  \mn@doi [\mnras] {10.1093/mnras/stw733}, \href
  {http://adsabs.harvard.edu/abs/2016MNRAS.459.2370T} {459, 2370}

\bibitem[\protect\citeauthoryear{{Walker}, {Mateo}, {Olszewski}, {Bailey},
  {Koposov}, {Belokurov}  \& {Evans}}{{Walker} et~al.}{2015}]{Walker2015}
{Walker} M.~G.,  {Mateo} M.,  {Olszewski} E.~W.,  {Bailey} III J.~I.,
  {Koposov} S.~E.,  {Belokurov} V.,   {Evans} N.~W.,  2015, \mn@doi [\apj]
  {10.1088/0004-637X/808/2/108}, \href
  {http://adsabs.harvard.edu/abs/2015ApJ...808..108W} {808, 108}

\bibitem[\protect\citeauthoryear{{Walker} et~al.,}{{Walker}
  et~al.}{2016}]{Walker2016}
{Walker} M.~G.,  et~al., 2016, \mn@doi [\apj] {10.3847/0004-637X/819/1/53},
  \href {http://adsabs.harvard.edu/abs/2016ApJ...819...53W} {819, 53}

\bibitem[\protect\citeauthoryear{{White} \& {Rees}}{{White} \&
  {Rees}}{1978}]{White1978}
{White} S.~D.~M.,  {Rees} M.~J.,  1978, \mn@doi [\mnras]
  {10.1093/mnras/183.3.341}, \href
  {http://adsabs.harvard.edu/abs/1978MNRAS.183..341W} {183, 341}

\bibitem[\protect\citeauthoryear{{Willman} \& {Strader}}{{Willman} \&
  {Strader}}{2012}]{Willman2012}
{Willman} B.,  {Strader} J.,  2012, \mn@doi [\aj] {10.1088/0004-6256/144/3/76},
  \href {http://adsabs.harvard.edu/abs/2012AJ....144...76W} {144, 76}

\bibitem[\protect\citeauthoryear{{York}, {Adelman}, {Anderson}, {Anderson}  \&
  {et al.}}{{York} et~al.}{2000}]{York2000}
{York} D.~G.,  {Adelman} J.,  {Anderson} Jr. J.~E.,  {Anderson} S.~F.,   {et
  al.} 2000, \mn@doi [\aj] {10.1086/301513}, \href
  {http://adsabs.harvard.edu/abs/2000AJ....120.1579Y} {120, 1579}

\bibitem[\protect\citeauthoryear{{Zinn}}{{Zinn}}{1985}]{Zinn1985}
{Zinn} R.,  1985, \mn@doi [\apj] {10.1086/163249}, \href
  {http://adsabs.harvard.edu/abs/1985ApJ...293..424Z} {293, 424}

\bibitem[\protect\citeauthoryear{{Zinn}}{{Zinn}}{1993}]{Zinn1993}
{Zinn} R.,  1993, in {Smith} G.~H.,  {Brodie} J.~P.,  eds,  Astronomical
  Society of the Pacific Conference Series Vol. 48, The Globular Cluster-Galaxy
  Connection. p.~38

\makeatother
\end{thebibliography}

{\small\it\noindent
$^1$Instituto de F\'\i sica, UFRGS, Caixa Postal 15051, Porto Alegre, RS - 91501-970, Brazil\\
$^2$Laborat\'orio Interinstitucional de e-Astronomia - LIneA, Rua Gal. Jos\'e Cristino 77, Rio de Janeiro, RJ - 20921-400, Brazil\\
$^3$George P. and Cynthia Woods Mitchell Institute for Fundamental Physics and Astronomy, and Department of Physics and Astronomy, Texas A\&M University, College Station, TX 77843, USA\\
$^4$Fermi National Accelerator Laboratory, P. O. Box 500, Batavia, IL 60510, USA\\
$^5$Kavli Institute for Cosmological Physics, University of Chicago, Chicago, IL 60637, USA\\
$^6$Department of Physics, University of Surrey, Guildford GU2 7XH, UK\\
$^7$Observat\'orio Nacional, Rua Gal. Jos\'e Cristino 77, Rio de Janeiro, RJ 20921-400, Brazil \\
$^8$Cerro Tololo Inter-American Observatory, National Optical Astronomy Observatory, Casilla 603, La Serena, Chile\\
$^{9}$Department of Physics \& Astronomy, University College London, Gower Street, London, WC1E 6BT, UK\\
$^{10}$Department of Physics and Electronics, Rhodes University, PO Box 94, Grahamstown, 6140, South Africa\\
$^{11}$LSST, 933 North Cherry Avenue, Tucson, AZ 85721, USA\\
$^{12}$CNRS, UMR 7095, Institut d'Astrophysique de Paris, F-75014, Paris, France\\
$^{13}$Sorbonne Universit\'es, UPMC Univ Paris 06, UMR 7095, Institut d'Astrophysique de Paris, F-75014, Paris, France\\
$^{14}$Department of Astronomy, University of Illinois, 1002 W. Green Street, Urbana, IL 61801, USA\\
$^{15}$National Center for Supercomputing Applications, 1205 West Clark St., Urbana, IL 61801, USA\\
$^{16}$Institut de F\'{\i}sica d'Altes Energies (IFAE), The Barcelona Institute of Science and Technology, Campus UAB, 08193 Bellaterra (Barcelona) Spain\\
$^{17}$Institute of Space Sciences, IEEC-CSIC, Campus UAB, Carrer de Can Magrans, s/n,  08193 Barcelona, Spain\\
$^{18}$Kavli Institute for Particle Astrophysics \& Cosmology, P. O. Box 2450, Stanford University, Stanford, CA 94305, USA\\
$^{19}$Department of Physics, California Institute of Technology, Pasadena, CA 91125, USA\\
$^{20}$Jet Propulsion Laboratory, California Institute of Technology, 4800 Oak Grove Dr., Pasadena, CA 91109, USA\\
$^{21}$Instituto de Fisica Teorica UAM/CSIC, Universidad Autonoma de Madrid, 28049 Madrid, Spain\\
$^{22}$Department of Astronomy, University of Michigan, Ann Arbor, MI 48109, USA\\
$^{23}$Department of Physics, University of Michigan, Ann Arbor, MI 48109, USA\\
$^{24}$SLAC National Accelerator Laboratory, Menlo Park, CA 94025, USA\\
$^{25}$Center for Cosmology and Astro-Particle Physics, The Ohio State University, Columbus, OH 43210, USA\\
$^{26}$Department of Physics, The Ohio State University, Columbus, OH 43210, USA\\
$^{27}$Astronomy Department, University of Washington, Box 351580, Seattle, WA 98195, USA\\
$^{28}$Australian Astronomical Observatory, North Ryde, NSW 2113, Australia\\
$^{29}$Instituci\'o Catalana de Recerca i Estudis Avan\c{c}ats, E-08010 Barcelona, Spain\\
$^{30}$Institute of Cosmology \& Gravitation, University of Portsmouth, Portsmouth, PO1 3FX, UK\\
$^{31}$Centro de Investigaciones Energ\'eticas, Medioambientales y Tecnol\'ogicas (CIEMAT), Madrid, Spain\\
$^{32}$School of Physics and Astronomy, University of Southampton,  Southampton, SO17 1BJ, UK\\
$^{33}$Instituto de F\'isica Gleb Wataghin, Universidade Estadual de Campinas, 13083-859, Campinas, SP, Brazil\\
$^{34}$Computer Science and Mathematics Division, Oak Ridge National Laboratory, Oak Ridge, TN 37831\\
}

\bsp	
\label{lastpage}
\end{document}